\documentclass[journal]{IEEEtran}
\usepackage{amsmath,amssymb,amsfonts}
\usepackage{array}
\usepackage{booktabs}
\usepackage{multirow}
\usepackage{makecell}
\usepackage{graphicx}
\usepackage[table]{xcolor}
\usepackage{textcomp}
\usepackage{stfloats}
\usepackage{url}
\usepackage{cite}
\graphicspath{{./}}
\newcommand{\citep}[1]{\cite{#1}}
\newcommand{\latentdim}{C_{\mathrm{lat}}}
\usepackage[ruled,vlined,linesnumbered]{algorithm2e}
\SetKwInput{KwIn}{Input}
\SetKwInput{KwOut}{Output}
\SetKw{Return}{return}
\definecolor{algorithmcomment}{RGB}{158, 200, 185}

\SetCommentSty{algorithmcommentstyle}
\SetKwComment{tcp}{\textcolor{algorithmcomment}{$\triangleright$\ }}{}
\newcommand{\linecomment}[1]{\tcp*[f]{#1}}
\SetAlgoNlRelativeSize{-1}
\newlength{\narrowfloatwidth}
\newlength{\widetabcolsep}
\AtBeginDocument{%
  \setlength{\narrowfloatwidth}{\columnwidth}%
  \setlength{\widetabcolsep}{4pt}}
\usepackage{placeins}
\usepackage[hidelinks]{hyperref}
\newcommand{\suppref}[1]{\ref{supp:#1}}
\newcommand{\mainref}[1]{\ref{#1}}
\begin{document}

\title{TOP: Transmission-Oriented Remote Sensing Pansharpening}

\author{Zi-Han~Cao,
        Jing-Ze~Liang,
        Shao-Jie~Wu,
        Jin-Cheng~Di,
        Liang-Jian~Deng,~\IEEEmembership{Senior~Member,~IEEE},
        and~Gemine~Vivone,~\IEEEmembership{Senior~Member,~IEEE}%
\thanks{This work was supported in part by the National Natural Science Foundation of China (NSFC) under Grant 12671624.}%
\thanks{Zi-Han Cao and Jing-Ze Liang are co-first authors and contributed equally to this work. Corresponding author: Liang-Jian Deng.}%
\thanks{Z.-H. Cao, J.-Z. Liang, S.-J. Wu, and J.-C. Di are with the University of Electronic Science and Technology of China, Chengdu, Sichuan 611731, China (e-mail: iamzihan666@gmail.com; liangjingze7@gmail.com; 2025110902003@std.uestc.edu.cn; 2025110901024@std.uestc.edu.cn).}%
\thanks{L.-J. Deng is with the University of Electronic Science and Technology of China, Chengdu, Sichuan 611731, China, and also with the Multi-Hazard Early Warning Key Laboratory of Sichuan Province, Chengdu, Sichuan 611731, China (e-mail: liangjian.deng@uestc.edu.cn).}%
\thanks{G. Vivone is with the National Research Council of Italy, Institute of Integrated Methodologies for Earth Observation (CNR-IMIOT), 85050 Tito, Italy (e-mail: gemine.vivone@imaa.cnr.it).}}

\markboth{TOP: Transmission-Oriented Remote Sensing Pansharpening}%
{Cao \MakeLowercase{\textit{et al.}}: TOP}

\maketitle

\begin{abstract}
Pansharpening has been extensively studied as a fundamental image fusion task in remote sensing. However, most existing methods focus on improving fusion quality under the implicit assumption that high-resolution observations have already been fully downlinked to the ground. As a result, existing models are highly vulnerable to perturbations introduced by lossy compression used for downlink transmission. 
To address this limitation, we present, to the best of our knowledge, the first transmission-oriented pansharpening (TOP) framework, which unifies compact latent coding, onboard pansharpening, and efficient downlink transmission within a single satellite-to-ground pipeline.
TOP encodes remote-sensing observations into a compact binary latent space, where a lightweight onboard fusion module can efficiently predict the fused binary latent code. Since only the compact fused code is downlinked, the transmission payload is substantially reduced. Overall, TOP enables low-payload downlink and lightweight onboard processing while maintaining high fusion quality under severe bandwidth constraints.
Experiments across multispectral and hyperspectral settings show that TOP consistently achieves superior fusion quality at higher compression ratios than representative pansharpening methods paired with standard lossy codecs. For example, on GaoFen-2, TOP attains a fusion PSNR of 31.4 dB at a downlink compression ratio of 50$\times$. This design also points toward transmission-oriented processing for other remote-sensing tasks. The code will be made available at \url{https://github.com/JingZe-Liang/TOP.git}.

\end{abstract}


\section{Introduction}\label{sec:introduction}

\IEEEPARstart{P}{ansharpening} reconstructs a high-spatial-resolution multispectral or hyperspectral image (HR-MSI/HR-HSI) by combining the spatial structures of a panchromatic (PAN) observation with the spectral information of a low-spatial-resolution multispectral or hyperspectral image (LR-MSI/LR-HSI), and underpins remote-sensing image interpretation~\citep{Deng2018VariationalPan,Wu2023LRTCFPan}. From the explicit detail-injection rules of component-substitution and multi-resolution-analysis methods~\citep{Laben2000GS,Garzelli2008BDSD,Aiazzi2002GLP} to the nonlinear mappings learned by deep networks~\citep{Deng2021FusionNet,Peng2023U2Net,Huang2025WFANet}, these methods share an implicit assumption: \textit{the input observations have already been fully delivered to the ground station and are directly available to the fusion model.}

\begin{figure*}[t]
    \centering
    \includegraphics[width=\textwidth]{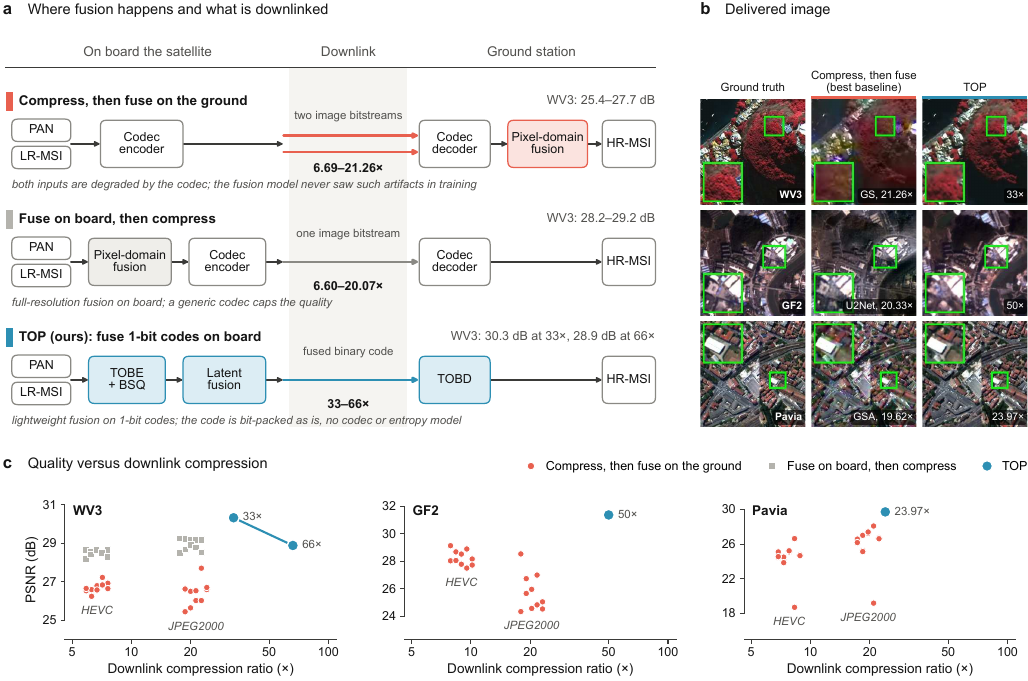}
    \caption{\textbf{TOP compared with existing deployment paths for pansharpening.} (a) Three ways of delivering a pansharpened image to the ground, with compression ratios and PSNR ranges from WorldView-3 (WV3). (b) Ground truth, the best compress-then-fuse baseline under JPEG2000, and TOP on WV3, GaoFen-2 (GF2), and Pavia Center. (c) PSNR versus downlink compression ratio for methods under HEVC or JPEG2000; onboard fusion followed by compression is evaluated on WV3 only (Sec.~\suppref{sec:supp_onboard_fusion} of the Supplementary Material).}
    \label{fig:intro_top_overview}
\end{figure*}

In practice, the PAN and LR-MSI/LR-HSI observations are acquired on board and must pass through bandwidth-limited downlink channels~\citep{Xiang2024TaskOrientedCompression}. A conventional deployment therefore compresses and transmits them independently and fuses the decompressed images on the ground (first row of Fig.~\ref{fig:intro_top_overview}(a)). Under restrictive bitrate budgets, lossy compression introduces spatial blurring, spectral distortion, and quantization artifacts, whereas pansharpening models are commonly trained on clean observations~\citep{Deng2021FusionNet,Peng2023U2Net,Huang2025WFANet}. This training--deployment mismatch can severely degrade the fusion result (Fig.~\ref{fig:intro_top_overview}(b)), and the codec-based baselines in Sec.~\ref{sec:exp_main_results} indeed degrade sharply under constrained downlink budgets. Recent work on degradation-robust fusion~\citep{Lv2026STAFuse} further highlights the importance of realistic input perturbations.

Fusing on board (second row of Fig.~\ref{fig:intro_top_overview}(a)) does not resolve this issue. The satellite must then generate a full HR-MSI/HR-HSI, which is computationally expensive on resource-constrained platforms, and the fused image must still be compressed with a standard codec such as JPEG2000~\citep{Taubman2002JPEG2000} or High Efficiency Video Coding (HEVC)~\citep{Sullivan2012HEVC}, whose rate--distortion performance then bounds the final quality~\citep{Balle2018ScaleHyperprior,Xiang2024TaskOrientedCompression}.

These limitations motivate a different question: \textit{rather than deciding whether to compress images before or after pansharpening, can transmission and pansharpening, which have so far been handled separately, benefit from each other?} To this end, we propose, to the best of our knowledge, the first transmission-oriented pansharpening (TOP) framework, which jointly designs compact latent coding, onboard pansharpening, and downlink transmission. TOP maps the PAN and LR-MSI/LR-HSI observations into binary latent codes and employs a lightweight latent fusion module to predict the fused HR-MSI/HR-HSI code on board (third row of Fig.~\ref{fig:intro_top_overview}(a)). Only this code is downlinked; the ground station decodes it into the final HR-MSI/HR-HSI or archives it for later reconstruction. TOP thus shifts the communication target from source images to a compact fused code.

TOP is built on a transmission-oriented binary tokenizer (TOBT), which maps PAN, LR-MSI/LR-HSI, and HR-MSI/HR-HSI observations into a shared latent space through an adaptive embedding for different numbers of bands and a Swin-based encoder~\citep{Liu2021Swin}. Binary spherical quantization (BSQ)~\citep{Zhao2025BSQ} then maps each latent entry to one bit. Besides reducing the payload, this bottleneck turns continuous-latent regression into an easier per-entry binary classification for the latent fusion module (Sec.~\ref{sec:exp_ablation_bsq}). The binary code is transmitted directly, without an entropy model~\citep{Balle2018ScaleHyperprior} or range asymmetric numeral systems (rANS) coding~\citep{Duda2013ANS} (Sec.~\ref{sec:exp_ablation_entropy}), and bit-flip augmentation makes the ground-side decoder robust to latent prediction and transmission errors.

We compare TOP with representative pansharpening methods whose inputs are compressed by HEVC or JPEG2000 before ground-side fusion, on the WorldView-3 (WV3) and GaoFen-2 (GF2) multispectral datasets and the Pavia Center hyperspectral dataset (hereafter Pavia). On all three datasets, TOP delivers a higher PSNR than every compressed baseline while operating at a higher compression ratio (Fig.~\ref{fig:intro_top_overview}(c)); on WV3, for example, it reaches 30.32 dB at $33\times$, whereas the strongest baseline reaches 27.70 dB at a lower ratio. At these ratios, 1~TB of source observations corresponds to roughly 15.2~GB of downlinked data on WV3 (at $66\times$), 20.0~GB on GF2, and 41.7~GB on Pavia. Our contributions are summarized as follows:

\begin{itemize}
\item We firstly introduce and formalize transmission-oriented pansharpening, a system-level design that jointly considers compact binary coding, onboard fusion, downlink communication, and ground-side reconstruction instead of assuming fully available lossless observations.

\item We build TOP, which performs pansharpening in a compact binary latent space and transmits only the fused HR-MSI/HR-HSI code; its BSQ bottleneck yields a code that is reconstructable, directly transmissible, and predictable for the latent fusion module.

\item We conduct experiments on multispectral and hyperspectral datasets, showing that TOP achieves superior fusion quality at substantially higher compression ratios than representative pansharpening methods paired with codecs.
\end{itemize}

\section[Related Work]{Related Work}\label{sec:related_work}

\subsection{Pansharpening}

Pansharpening fuses the spatial details of a PAN image with the spectral information of an LR-MSI/LR-HSI image~\citep{Deng2018VariationalPan,Wu2023LRTCFPan} and belongs to the broader family of image fusion tasks, which also includes infrared--visible, multi-exposure, and multi-focus fusion~\citep{Li2025UMCFuse,Luo2023MEF,Li2020DRPL}. Traditional methods are based on component substitution, such as GS~\citep{Laben2000GS}, BDSD~\citep{Garzelli2008BDSD}, GSA~\citep{Aiazzi2007GSA}, and PRACS~\citep{Choi2011PRACS}; multi-resolution analysis, such as GLP~\citep{Aiazzi2002GLP}, AWLP~\citep{Otazu2005AWLP}, and morphological filters~\citep{Restaino2016MF}; or variational, Bayesian, and low-rank formulations~\citep{Deng2018VariationalPan,Wang2019BayesianPan,Lu2022UnifiedPan,Wu2023LRTCFPan}. They are efficient and interpretable, but predefined models may cause spatial or spectral distortions. Deep multispectral models have introduced detail injection~\citep{Deng2021FusionNet}, dual-branch spatial--spectral modeling~\citep{Peng2023U2Net}, frequency-aware or adaptive feature extraction~\citep{Huang2025WFANet,Huang2025ADWM,Wang2025ARConv}, degradation-aware modeling~\citep{Xu2026DTEA}, and zero-shot adaptation~\citep{Cao2024ZSPan}. Hyperspectral pansharpening must further handle high spectral dimensionality and sensor variability, and relies on Transformer-based fusion and equivariant imaging~\citep{Bandara2022HyperTransformer,Wang2026EquivariantHSI}, image-specific priors and band-wise propagation~\citep{Bandara2022DIPHyperKite,Guarino2024RPNN,Guarino2025RhoPNN}, and diffusion-based zero-shot guidance~\citep{Jiang2025uTDSP,Xiao2025DMZS}. Degradation-robust fusion methods~\citep{Lv2026STAFuse} handle generic low-quality inputs rather than downlink compression artifacts. All these methods assume that the source observations are fully available at the processing platform, whereas TOP performs pansharpening before downlink in a compact binary latent space.

\subsection{Remote-Sensing Image Compression and Transmission}
The growing volume of remote-sensing imagery strains onboard storage and satellite-to-ground communication~\citep{Xiang2024TaskOrientedCompression}. Standard codecs such as JPEG2000~\citep{Taubman2002JPEG2000} and HEVC~\citep{Sullivan2012HEVC} reduce payloads through transform coding, quantization, and entropy coding. Lossless and near-lossless remote-sensing codecs such as CCSDS 123.0-B-2~\citep{ccsds} target low-complexity, distortion-bounded compression rather than the high-compression regime studied here, and are therefore not included in our comparison. Neural image compression instead learns nonlinear transforms with quantized latents under rate--distortion optimization~\citep{Balle2018ScaleHyperprior}. Existing pipelines combine compression and pansharpening in one of two orders, either compressing the source observations and fusing on the ground or fusing on board and compressing the fused image, and in both cases treat fusion and compression as separate stages. TOP instead fuses within a compact binary latent space and downlinks the fused code as the payload.

\subsection{Discrete Visual Representation and Quantization}

Transmitting a learned latent requires discretizing it. VQ-VAE~\citep{VanDenOord2017VQVAE} assigns latent vectors to a learned codebook whose capacity and utilization are hard to control. Finite scalar quantization~\citep{Mentzer2024FSQ} maps each dimension to fixed scalar levels, lookup-free quantization~\citep{Yu2024MAGVITv2} derives token indices directly from binary states, and BSQ~\citep{Zhao2025BSQ} projects latents onto a unit hypersphere before sign quantization. Such binary codes are highly compact and thus well suited to transmission, so TOP adopts BSQ as a shared latent code for all three modalities. Unlike generative models, which predict discrete tokens with autoregressive or masked-token objectives~\citep{Yu2024MAGVITv2}, TOP predicts the target binary code conditionally from the PAN and LR-MSI/LR-HSI codes, and this prediction serves as both the downlink payload and the input to the ground-side decoder.

\begin{algorithm}[!t]
\caption{Deployment-time inference of TOP.}
\label{alg:top_inference}
\footnotesize
\DontPrintSemicolon
\KwIn{PAN observation $\mathbf{P}$; upsampled LR-MSI/LR-HSI observation $\mathbf{M}$; trained encoder $\mathcal{E}_{\theta}$; quantizer $\mathcal{Q}$; latent fusion module $\mathcal{F}_{\psi}$; ground-side decoder $\mathcal{D}_{\phi}$}
\KwOut{Reconstructed HR-MSI/HR-HSI image $\widehat{\mathbf{H}}$}
\BlankLine
\tcp{\textbf{Onboard Encoding, Fusion, and Downlink}}

$\mathbf{B}_{\mathrm{P}} \gets \mathcal{Q}(\mathcal{E}_{\theta}(\mathbf{P}))$ \linecomment{Encode PAN}\\
$\mathbf{B}_{\mathrm{M}} \gets \mathcal{Q}(\mathcal{E}_{\theta}(\mathbf{M}))$ \linecomment{Encode spectral image}\\

$\mathbf{O} \gets \mathcal{F}_{\psi}\!\left(\operatorname{Concat}_{c}[\mathbf{B}_{\mathrm{M}},\mathbf{B}_{\mathrm{P}}]\right)$ \linecomment{Predict binary-state logits}\\
$\widehat{\mathbf{B}}_{\mathrm{H}} \gets \bigl(2\arg\,\max\limits_{a\in\{0,1\}}\mathbf{O}_{a}-\mathbf{1}\bigr)/\sqrt{\latentdim}$ \linecomment{Normalize the fused code}\\
$\mathbf{S}_{\mathrm{down}} \gets \operatorname{BitPack}(\widehat{\mathbf{B}}_{\mathrm{H}})$ \linecomment{Pack the binary payload}\\
Downlink $\mathbf{S}_{\mathrm{down}}$ to the ground station\;
\BlankLine
\tcp{\textbf{Ground-Side Reconstruction}}
$\widetilde{\mathbf{B}}_{\mathrm{H}} \gets \operatorname{BitUnpack}(\mathbf{S}_{\mathrm{down}})$ \linecomment{Recover the received fused code}\\

$\widehat{\mathbf{H}} \gets \mathcal{D}_{\phi}(\widetilde{\mathbf{B}}_{\mathrm{H}})$ \linecomment{Decode the fused HR image}\\
\Return $\widehat{\mathbf{H}}$\;
\end{algorithm}
\section{Methodology}\label{sec:method}
Sec.~\ref{secoverview} introduces the notation and the deployment-time pipeline. We then detail the components of TOBT, namely the transmission-oriented binary encoder (TOBE; Sec.~\ref{sec:method_tokenization}), BSQ (Sec.~\ref{Sec BSQ}), and the transmission-oriented binary decoder (TOBD; Sec.~\ref{sec:method_tobd_decoder}), together with the latent fusion module (Sec.~\ref{sec:method_latent_fusion}), the training strategy (Sec.~\ref{sec:method_objective}), and the overall formulation of TOP (Sec.~\ref{formulation}). Implementation details are provided in Sec.~\suppref{sec:supp_implementation} of the Supplementary Material.

\subsection{System Overview}
\label{secoverview}
\begin{figure*}[!t]
\centering
\includegraphics[width=\textwidth]{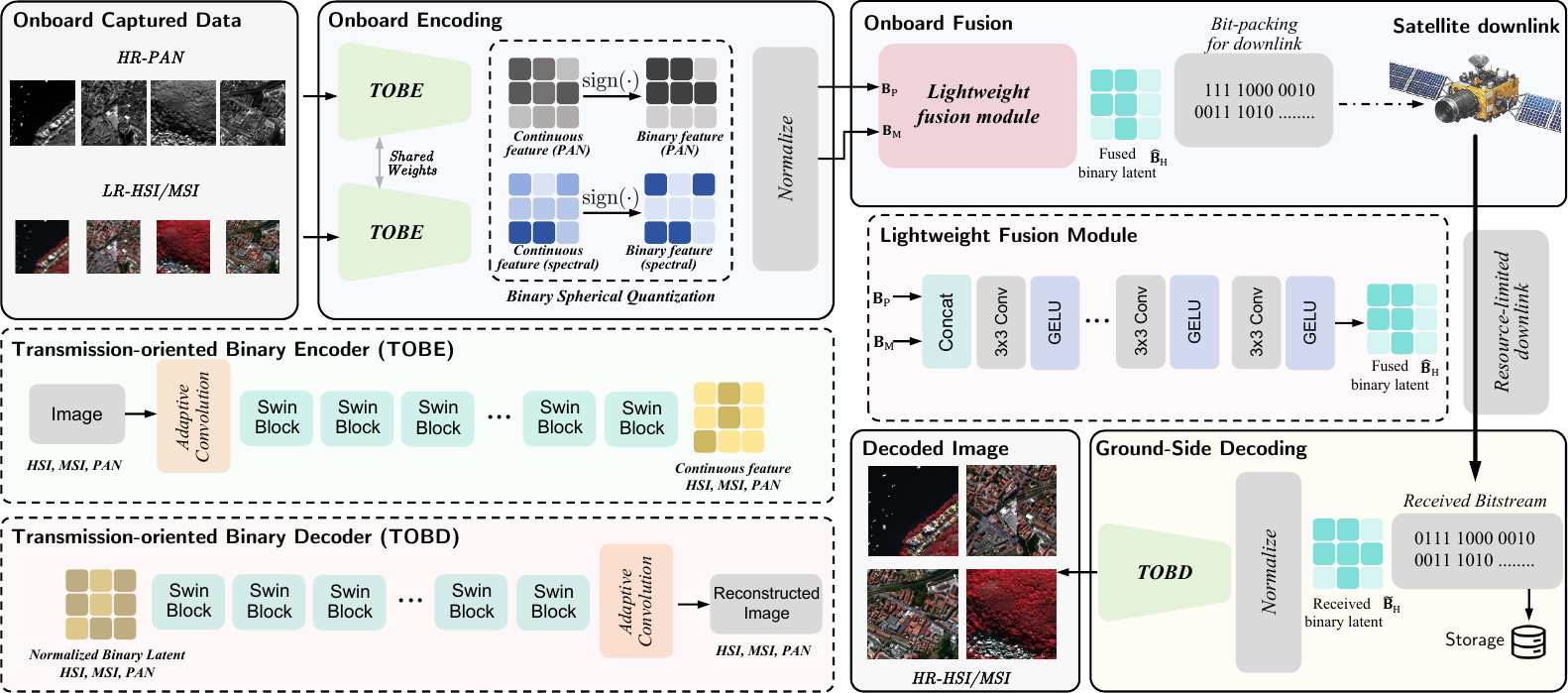}
\caption{\textbf{Overall pipeline of TOP.} On board, the observations are encoded by TOBE, binarized, and fused; the fused code is bit-packed, downlinked, and decoded by TOBD or stored on the ground. Downsampling is omitted for clarity; configurations are listed in Tab.~\suppref{tab:implementation_architecture} of the Supplementary Material. GELU: Gaussian error linear unit.}
\label{fig:method_overview}
\end{figure*}

Let $\mathbf{P}\in\mathbb{R}^{1\times H\times W}$, $\mathbf{M}\in\mathbb{R}^{C\times H\times W}$, and $\mathbf{H}\in\mathbb{R}^{C\times H\times W}$ denote the PAN observation, the upsampled LR-MSI/LR-HSI observation, and the target HR-MSI/HR-HSI image, respectively. During deployment, TOP divides pansharpening into onboard latent fusion and ground-side reconstruction (Fig.~\ref{fig:method_overview}). On board, the shared TOBE $\mathcal{E}_{\theta}$ and the BSQ quantizer $\mathcal{Q}$ encode $\mathbf{P}$ and $\mathbf{M}$ into normalized binary latent codes $\mathbf{B}_{\mathrm{P}}$ and $\mathbf{B}_{\mathrm{M}}$, which are concatenated along the channel dimension and processed by the lightweight latent fusion module $\mathcal{F}_{\psi}$ to predict the fused code $\widehat{\mathbf{B}}_{\mathrm{H}}$. During training, this prediction is supervised by $\mathbf{B}_{\mathrm{H}}$, the code obtained by encoding $\mathbf{H}$; during inference, $\widehat{\mathbf{B}}_{\mathrm{H}}$ is directly bit-packed and transmitted. At the ground station, the received code is unpacked as $\widetilde{\mathbf{B}}_{\mathrm{H}}$, and the decoder $\mathcal{D}_{\phi}$ reconstructs the fused image $\widehat{\mathbf{H}}\in\mathbb{R}^{C\times H\times W}$, which approximates the ground-truth (GT) image $\mathbf{H}$. Algorithm~\ref{alg:top_inference} summarizes the deployment-time inference. The variables $\mathbf{H}$ and $\mathbf{B}_{\mathrm{H}}$ define only the training target; neither is available on board during deployment. The fused code is inferred solely from $\mathbf{P}$ and $\mathbf{M}$, so the payload is a predicted fused representation rather than a compressed copy of a pre-existing HR-MSI/HR-HSI.

\subsection{Transmission-Oriented Binary Encoder}
\label{sec:method_tokenization}
TOBE maps observations with different numbers of spectral bands into a shared, compact latent feature space before binary quantization (middle-left panel of Fig.~\ref{fig:method_overview}).

\subsubsection{Band-Adaptive Embedding}
TOBE embeds all modalities with an adaptive convolution that maintains a learnable base kernel $\mathbf{W}_0$ with $C_0$ input channels. For an incoming modality with $C_\mathrm{in}$ spectral bands, the kernel is resampled along its input-channel dimension:
\begin{equation}
\mathbf{W}_{\mathrm{input}}^{(c)}
=
\sum_{j=1}^{C_0}
\alpha_{c,j}\mathbf{W}_{0}^{(j)},
\end{equation}
where the superscripts $(c)$ and $(j)$ select input-channel slices of the adapted and base kernels, $c=1,\ldots,C_{\mathrm{in}}$, and $\alpha_{c,j}$ is the bicubic resampling weight from base channel $j$ to adapted channel $c$. PAN and spectral inputs are thus mapped to the same feature dimension with shared weights. TOBD applies the same adaptation along the output-channel dimension.

\subsubsection{Swin-Based Multi-Stage Feature Extraction}
The embedded features are processed by a Swin Transformer encoder~\citep{Liu2021Swin}, whose windowed self-attention provides stronger content-adaptive spatial--spectral modeling than CNNs at a much lower cost than global self-attention~\citep{Dosovitskiy2021ViT} (Tab.~\ref{tab:tokenizer_complexity}); the ablation in Sec.~\ref{sec:exp_ablation_arch} verifies this choice. With $\mathbf{F}_0$ denoting the embedded feature, the $L$ encoder stages compute
\begin{equation}
\mathbf{F}_{\ell}
=
\mathcal{S}_{\ell}
\left(
\mathcal{R}_{\ell}
\left(
\mathbf{F}_{\ell-1}
\right)
\right),
\qquad
\ell=1,\ldots,L,
\end{equation}
where $\mathcal{R}_{\ell}$ is a downsampling operator, applied only until the target factor $s$ is reached, and $\mathcal{S}_{\ell}$ is a stack of Swin blocks whose depth varies with $\ell$ (Tab.~\suppref{tab:implementation_architecture} of the Supplementary Material). The output $\mathbf{F}_L$ has $\latentdim$ channels and is spatially reduced by a factor of $s$ along each dimension.

\begin{table}[!t]
\caption{Complexity of Candidate TOBE Backbone Blocks ($N_{\mathrm{tok}}$ Tokens, Width $d$, Kernel Size $k$, Window Size $M$)}
\label{tab:tokenizer_complexity}
\footnotesize
\centering
\begin{tabular*}{\narrowfloatwidth}{@{\extracolsep{\fill}}lll@{}}
\toprule
Block & Modeling behavior & Main complexity $\downarrow$ \\
\midrule
CNN & Fixed, local & $\mathcal{O}(N_{\mathrm{tok}}k^2d^2)$ \\
Global attention & Adaptive, all tokens & $\mathcal{O}(N_{\mathrm{tok}}^2d+N_{\mathrm{tok}}d^2)$ \\
Swin & Adaptive, windowed & $\mathcal{O}(N_{\mathrm{tok}}M^2d+N_{\mathrm{tok}}d^2)$ \\
\bottomrule
\end{tabular*}
\end{table}

\subsection{Binary Spherical Quantization}
\label{Sec BSQ}

Although spatially compressed, $\mathbf{F}_{L}$ is still continuous-valued. TOBT therefore adopts BSQ~\citep{Zhao2025BSQ}, a binary tokenization technique from generative visual representation learning that maps the latent vector at each location to a binary code on the unit hypersphere. BSQ reduces the payload to one bit per latent entry and provides the latent fusion module with a compact and explicit discrete space that is well suited to fusion (Sec.~\ref{sec:exp_ablation_bsq}). Let $\mathbf{f}_{i}\in\mathbb{R}^{\latentdim}$ denote the channel vector of $\mathbf{F}_{L}$ at latent location $i=1,\ldots,N$, with $N=HW/s^{2}$. BSQ performs deterministic sign quantization
\begin{equation}
\mathbf{b}_{i}
=
\operatorname{sign}\left({\mathbf{f}}_{i}\right)\in\{-1,+1\}^{\latentdim},
\end{equation}
where $\operatorname{sign}(\cdot)$ acts elementwise, maps zero to $+1$, and is trained with a straight-through estimator. Collecting the codes over all locations gives the normalized binary latent code
\begin{equation}
\begin{aligned}
\mathbf{B}
&=\frac{1}{\sqrt{\latentdim}}
\operatorname{reshape}~\!\Bigl(\bigl[\mathbf{b}_1,\ldots,\mathbf{b}_N\bigr]\Bigr)\\
&\in\left\{-\frac{1}{\sqrt{\latentdim}},+\frac{1}{\sqrt{\latentdim}}\right\}^{\latentdim\times H/s\times W/s},
\end{aligned}
\end{equation}
where `$\operatorname{reshape}(\cdot)$' arranges the $\latentdim\times N$ matrix of column vectors into a tensor in a fixed raster order, and $\mathbf{B}$ denotes the code of any of the three modalities. During training, soft assignment and grouped entropy regularization encourage effective code utilization~\citep{Zhao2025BSQ}. In practice, the latent channels are split into several groups and the entropy terms are computed per group; for simplicity, the main text treats all channels as a single group. Its details are given in Sec.~\suppref{sec:supp_bsq} of the Supplementary Material.

\subsection{Latent Fusion Module}
\label{sec:method_latent_fusion}

From the concatenated codes $\mathbf{B}_{\mathrm{M}}$ and $\mathbf{B}_{\mathrm{P}}$, the latent fusion module $\mathcal{F}_{\psi}$ predicts two logits for each latent entry:
\begin{equation}
\label{eq:latent_logits}
\mathbf{O}
=
\mathcal{F}_{\psi}
\left(
\operatorname{Concat}_{c}
\left[
\mathbf{B}_{\mathrm{M}},
\mathbf{B}_{\mathrm{P}}
\right]
\right)
\in
\mathbb{R}^{2\times \latentdim\times H/s\times W/s}.
\end{equation}
and the fused code is obtained by selecting a binary state independently for each entry:
\begin{equation}
\label{Fusion}
\widehat{\mathbf{B}}_{\mathrm{H}}
=
\frac{1}{\sqrt{\latentdim}}\left(
2\,\arg\,\max\limits_{a\in\{0,1\}}\mathbf{O}_a
-\mathbf{1}\right),
\end{equation}
where $\mathbf{O}_a$ is the slice of $\mathbf{O}$ for state $a\in\{0,1\}$, the $\arg\max$ is taken elementwise, and $\mathbf{1}$ is an all-ones tensor of the latent shape.

\subsection{Binary Downlink and TOBD}
\label{sec:method_tobd_decoder}

The satellite downlinks the bitstream $\mathbf{S}_{\mathrm{down}}=\operatorname{BitPack}(\widehat{\mathbf{B}}_{\mathrm{H}})$, where `$\operatorname{BitPack}(\cdot)$' maps negative entries to 0 and nonnegative entries to 1 and serializes them in a fixed order, together with the latent raster shape, whose negligible size is omitted from the compression ratio in Eq.~\eqref{compression ratio}. Unlike the original BSQ, which further compresses the codes with an autoregressive prior and arithmetic coding~\citep{Zhao2025BSQ}, TOP uses no entropy model or rANS codec~\citep{Duda2013ANS}, as supported by Sec.~\ref{sec:exp_ablation_entropy}.

On the ground side, the received bitstream is unpacked and
decoded as
\begin{equation}
\label{eq:ground_decoding}
\begin{aligned}
\widetilde{\mathbf{B}}_{\mathrm{H}}
&=
\operatorname{BitUnpack}(\mathbf{S}_{\mathrm{down}}),\\
\widehat{\mathbf{H}}
&=
\mathcal{D}_{\phi}(\widetilde{\mathbf{B}}_{\mathrm{H}}),
\end{aligned}
\end{equation}
where `$\operatorname{BitUnpack}(\cdot)$' inverts `$\operatorname{BitPack}(\cdot)$' by mapping each bit $z\in\{0,1\}$ to $(2z-1)/\sqrt{\latentdim}$; the received code $\widetilde{\mathbf{B}}_{\mathrm{H}}$ may differ from $\widehat{\mathbf{B}}_{\mathrm{H}}$ if transmission errors occur. $\mathcal{D}_{\phi}$ follows the stage layout of $\mathcal{E}_{\theta}$ in reverse order, possibly with different numbers of Swin blocks per stage (lower-left panel of Fig.~\ref{fig:method_overview}), and outputs the final HR-MSI/HR-HSI reconstruction $\widehat{\mathbf{H}}$. The binary code can also be archived for deferred decoding, which yields considerable storage savings.

\begin{figure*}[!t]
\centering
\includegraphics[width=\textwidth]{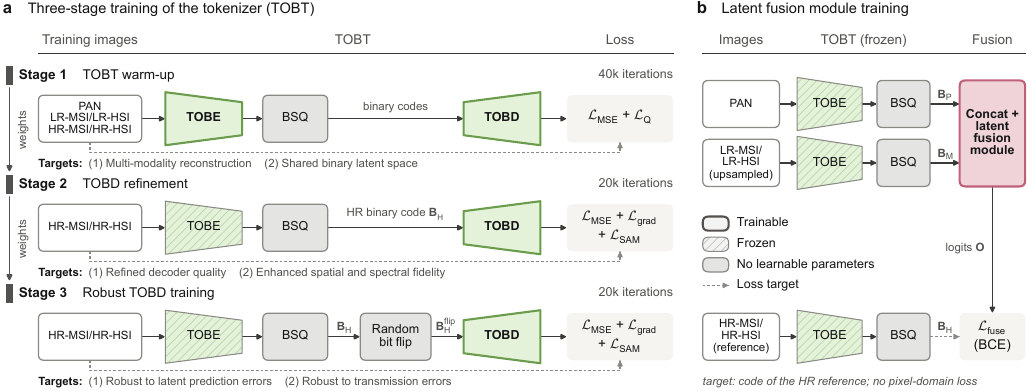}
\caption{\textbf{Training pipeline of the proposed method.} (a) Three-stage TOBT training: stage 1 trains TOBE and TOBD jointly (40k iterations); stages 2 and 3 refine TOBD with TOBE frozen (20k iterations each), stage 3 with random bit flips. (b) Latent fusion training with TOBT frozen. Bold outlines: trainable; hatched: frozen; gray: parameter-free; dashed arrows: loss targets.}
\label{fig:training_pipe}
\end{figure*}

\subsection{Optimization Objectives}\label{sec:method_objective}

We first train TOBT, including its TOBE and TOBD, and then train the latent fusion module with the frozen tokenizer (Fig.~\ref{fig:training_pipe}).

\subsubsection{Three-Stage TOBT Training}
\paragraph{Loss Components}

For an image-domain target $\mathbf{X}$ and its reconstruction $\widehat{\mathbf{X}}$, the mean squared error (MSE) and Sobel-gradient losses are
\begin{equation}
\mathcal{L}_{\mathrm{MSE}}
=
\frac{\|\widehat{\mathbf{X}}-\mathbf{X}\|_2^2}{CHW},
\qquad
\mathcal{L}_{\mathrm{grad}}
=
\frac{\|\nabla_{\mathrm{S}}\widehat{\mathbf{X}}-\nabla_{\mathrm{S}}\mathbf{X}\|_1}{CHW},
\end{equation}
where the norms are computed over vectorized entries and $\nabla_{\mathrm{S}}$ is the channel-wise Sobel operator; $\mathcal{L}_{\mathrm{grad}}$ penalizes edge discrepancies when refining the ground-side decoder. For spectral fidelity, the spectral angle mapper (SAM) loss~\citep{Kruse1993SAM} averages the angle, in radians, between the reconstructed and target spectral vectors $\widehat{\mathbf{x}}_i,\mathbf{x}_i\in\mathbb{R}^{C}$ over all spatial locations:
\begin{equation}
\mathcal{L}_{\mathrm{SAM}}
=
\frac{1}{HW}\sum_{i=1}^{HW}
\arccos
\Biggl(
\frac{\langle \widehat{\mathbf{x}}_i,\mathbf{x}_i\rangle}
{\|\widehat{\mathbf{x}}_i\|_2\|\mathbf{x}_i\|_2}
\Biggr).
\end{equation}
Unlike MSE, SAM is insensitive to uniform spectral scaling. The quantization loss follows BSQ~\citep{Zhao2025BSQ}: a commitment term pulls the normalized continuous latent toward its binary code, and an entropy term prevents the encoder from collapsing onto a few binary patterns:
\begin{equation}
\label{eq:quantization_loss}
\mathcal{L}_{\mathrm{Q}}
=
\beta\mathcal{L}_{\mathrm{commit}}
+
\mathcal{L}_{\mathrm{entropy}},
\end{equation}
where $\beta$ weights the commitment term (Sec.~\suppref{sec:supp_bsq} of the Supplementary Material).

\paragraph{Training Stages}

In stage~1 (Fig.~\ref{fig:training_pipe}(a)), TOBE and TOBD are jointly trained on all three modalities to stabilize the shared binary latent space before the decoder is refined, using
\begin{equation}
\mathcal{L}_{\mathrm{TOBT}}
=
\lambda_{\mathrm{MSE}}\mathcal{L}_{\mathrm{MSE}}
+
\lambda_{\mathrm{Q}}\mathcal{L}_{\mathrm{Q}},
\end{equation}
where $\lambda_{\mathrm{MSE}}$ and $\lambda_{\mathrm{Q}}$ are weighting coefficients. Stage~2 targets the full TOP setting, in which only the HR code is transmitted: TOBE is frozen, so the quantization loss is dropped, and TOBD is fine-tuned on the HR modality with
\begin{equation}
\mathcal{L}_{\mathrm{TOBD}}
=
\mathcal{L}_{\mathrm{MSE}}
+
\lambda_{\mathrm{grad}}\mathcal{L}_{\mathrm{grad}}
+
\lambda_{\mathrm{SAM}}\mathcal{L}_{\mathrm{SAM}},
\end{equation}
where the weights $\lambda_{\mathrm{grad}}$ and $\lambda_{\mathrm{SAM}}$ are listed in Sec.~\suppref{sec:supp_implementation} of the Supplementary Material.

\begin{figure}[!t]
\centering
\includegraphics[width=\narrowfloatwidth]{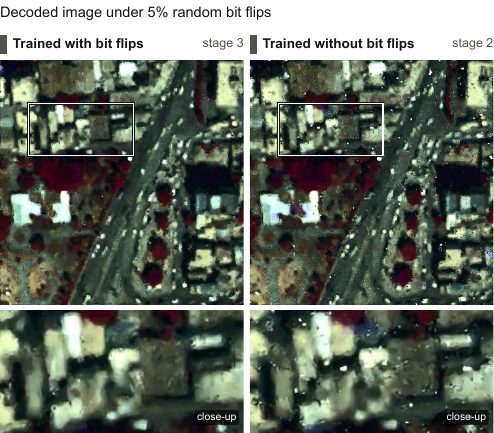}
\caption{\textbf{WV3 reduced-resolution MSI reconstruction under 5\% bit flips}.
Left: TOBD after stage 3 (with bit-flip augmentation). Right: TOBD after stage 2 (without). White boxes mark the close-ups below.}
\label{fig:ablation_bitflip_visual}
\end{figure}

Stage~3 copes with binary latent errors caused by imperfect latent fusion or by transmission. It keeps the stage-2 setting but injects random bit flips into the HR code before decoding:
\begin{equation}
\mathbf{B}_{\mathrm{H}}^{\mathrm{flip}}
=
(\mathbf{1}-2\mathbf{R})\odot\mathbf{B}_{\mathrm{H}},
\end{equation}
where the entries of the mask $\mathbf{R}\in\{0,1\}^{\latentdim\times H/s\times W/s}$ are independently drawn from $\operatorname{Bernoulli}(p(t))$, and an entry equal to 1 reverses the sign of the corresponding latent entry. The flip probability is linearly ramped from $p_{\mathrm{start}}=10^{-5}$ to $p_{\mathrm{end}}=3\times10^{-3}$ over the first 2000 iterations:
\begin{equation}
p(t)
=
p_{\mathrm{start}}
+
\min\left(\frac{t}{2000},1\right)
\left(p_{\mathrm{end}}-p_{\mathrm{start}}\right),
\end{equation}
where $t$ is the stage-3 iteration index. TOBD thereby learns to correct occasional bit errors by exploiting the redundancy of the code. Under 5\% bit flips (Fig.~\ref{fig:ablation_bitflip_visual}), the stage-3 decoder suppresses the speckle-like artifacts of the stage-2 decoder and better preserves spatial structures and colors; Sec.~\ref{sec:exp_ablation_robustness} quantifies this effect.

\subsubsection{Latent Fusion Module Training}

With all tokenizer parameters frozen, the PAN, upsampled LR-MSI/LR-HSI, and HR-MSI/HR-HSI images of each training sample are encoded into $\mathbf{B}_{\mathrm{P}}$, $\mathbf{B}_{\mathrm{M}}$, and $\mathbf{B}_{\mathrm{H}}$. The latent fusion module predicts the logits $\mathbf{O}$ from $\mathbf{B}_{\mathrm{M}}$ and $\mathbf{B}_{\mathrm{P}}$ as in Eq.~\eqref{eq:latent_logits}, and the fixed code $\mathbf{B}_{\mathrm{H}}$ supervises it, so the training target uses the same binary representation as the transmitted payload. The module is trained with a binary cross-entropy (BCE) loss on the logits:

\begin{equation}
\mathcal{L}_{\mathrm{fuse}} = f_{\mathrm{BCEWithLogits}}(\mathbf{O},\mathbf{T}),
\end{equation}
where the target $\mathbf{T}\in\{0,1\}^{2\times \latentdim\times H/s\times W/s}$ has the same shape as $\mathbf{O}$, with state slices $\mathbf{T}_1=\mathbb{I}[\mathbf{B}_{\mathrm{H}}>0]$ and $\mathbf{T}_0=\mathbf{1}-\mathbf{T}_1$ (elementwise), and the loss averages the scalar BCE terms over both states and all latent entries (Sec.~\suppref{app:bce_logits} of the Supplementary Material). No pixel-domain loss is back-propagated through TOBD during this training.

\subsection{Transmission-Oriented Pansharpening Formulation}
\label{formulation}

At the system level, TOP can be summarized as
\begin{equation}
\begin{aligned}
\min_{\theta,\psi,\phi}\quad
& \mathcal{L}_{\mathrm{fid}}
\left(\widehat{\mathbf{H}}, \mathbf{H}\right) \\
\text{s.t.}\quad
& \widehat{\mathbf{H}}
= \mathcal{D}_{\phi}
(\widetilde{\mathbf{B}}_{\mathrm{H}}),  |\mathbf{S}_{\mathrm{down}}|
\leq B_{\max},
\end{aligned}
\label{eq:top_formulation}
\end{equation}
where the downlinked bitstream $\mathbf{S}_{\mathrm{down}}$ and the received code $\widetilde{\mathbf{B}}_{\mathrm{H}}$ are obtained from $(\mathbf{P},\mathbf{M})$ through the encoding, latent-fusion, and bit-packing steps of Alg.~\ref{alg:top_inference} (Eqs.~\eqref{eq:latent_logits}--\eqref{eq:ground_decoding}), $\mathcal{L}_{\mathrm{fid}}(\cdot)$ collectively represents the fidelity objectives used throughout TOP, and $B_{\max}$ denotes the maximum allowable downlink budget.

In contrast, conventional pansharpening mainly considers
\begin{equation}
\min_{\omega}\quad
\mathcal{L}_{\mathrm{fus}}
\left(
\mathbf{H},
\mathcal{G}_{\omega}(\mathbf{P},\mathbf{M})
\right),
\label{eq:conventional_formulation}
\end{equation}
where $\mathcal{G}_{\omega}$ is a conventional pansharpening model with learnable parameters $\omega$ and $\mathcal{L}_{\mathrm{fus}}$ is its fusion objective. Unlike this formulation, TOP explicitly incorporates the downlink constraint into the pansharpening problem.

\section{Experiments}\label{sec:experiments}

\subsection{Experimental Settings}\label{sec:exp_settings}
\subsubsection{Datasets}
We use the WV3 (8 bands) and GF2 (4 bands) multispectral datasets from PanCollection~\citep{Deng2022MachineLearningPansharpening} and the 102-band Pavia Center hyperspectral dataset distributed through HyperPanCollection and used in Hyper-DSNet~\citep{zhuo2022deep}, all with a spatial scale factor of four. WV3, GF2, and Pavia contain 9,714, 19,809, and 1,512 training samples and 1,080, 2,201, and 168 validation samples, respectively; the validation samples are held-out training patches distinct from the test scenes. Training samples consist of $64\times64$ PAN, upsampled LR-MSI/LR-HSI, and HR-MSI/HR-HSI patches together with $16\times16$ native LR-MSI/LR-HSI patches. For WV3 and GF2, we test on 20 reduced-resolution scenes of size $256\times256$, simulated following Wald's protocol~\citep{Wald1997Protocol}, and on 20 full-resolution scenes of size $512\times512$ without GT. For Pavia, the \texttt{Full} test split contains two $400\times400$ scenes with HR-HSI references. Details are given in Sec.~\suppref{sec:supp_datasets} of the Supplementary Material.

\subsubsection{Compared Methods}
The multispectral baselines are GS~\citep{Laben2000GS}, BDSD~\citep{Garzelli2008BDSD}, GLP~\citep{Aiazzi2002GLP}, and MF~\citep{Restaino2016MF} (traditional); FusionNet~\citep{Deng2021FusionNet}, U2Net~\citep{Peng2023U2Net}, WFANet~\citep{Huang2025WFANet}, ADWM~\citep{Huang2025ADWM}, and ARConv~\citep{Wang2025ARConv} (supervised); and ZS-PAN~\citep{Cao2024ZSPan} (zero-shot). The hyperspectral baselines are AWLP~\citep{Otazu2005AWLP}, BDSD-PC~\citep{Vivone2019BDSDPC}, BT-H~\citep{Lolli2017HazeCorrection}, GSA~\citep{Aiazzi2007GSA}, MF~\citep{Restaino2016MF}, and PRACS~\citep{Choi2011PRACS} (traditional); HyperTransformer~\citep{Bandara2022HyperTransformer} (supervised); and DIP-HyperKite~\citep{Bandara2022DIPHyperKite}, R-PNN~\citep{Guarino2024RPNN}, uTDSP~\citep{Jiang2025uTDSP}, DM-ZS~\citep{Xiao2025DMZS}, and $\rho$-PNN~\citep{Guarino2025RhoPNN} (zero-shot).

\subsubsection{Transmission-Oriented Protocol and Communication Metrics}
\label{sec:transmission protocol}
All baselines follow the `compress first, fuse later' pipeline with JPEG2000~\citep{Taubman2002JPEG2000} (OpenJPEG, with a search over quality layers to match the target ratio) or HEVC~\citep{Sullivan2012HEVC} (libx265 with constant rate factor control to match the target ratio). The \textit{compression ratio} $R$ divides the raw sensor-side data volume by the size $B_{\mathrm{transmitted}}$ of the transmitted bitstream. To avoid inflating $R$, the raw volume uses the native, non-upsampled LR-MSI/LR-HSI:
\begin{equation}
R = \frac{bHW + bC\left(\frac{H}{r}\right)\left(\frac{W}{r}\right)}{B_{\mathrm{transmitted}}},
\label{compression ratio}
\end{equation}
where the PAN image has size $H\times W$, the LR-MSI/LR-HSI has $C$ bands and size $\frac{H}{r}\times\frac{W}{r}$, $r$ is the spatial scale factor, and $b$ is the native sensor bit depth (11 for WV3, 10 for GF2, and 13 for Pavia, based on the official specifications). For a codec-based baseline, $B_{\mathrm{transmitted}}$ is the total size of the separately encoded PAN and native LR-MSI/LR-HSI streams; for TOP, it is the size of the single bit-packed fused latent. Using the native spectral image in the numerator is essential because interpolation creates samples without adding sensor information and would otherwise overstate the apparent compression ratio. The \textit{bits per pixel per band} (bppb) normalizes the payload by the size of the target HR-MSI/HR-HSI:
\begin{equation}
\mathrm{bppb}
=
\frac{B_{\mathrm{transmitted}}}{C H W}.
\label{bppb}
\end{equation}

\begin{figure}[!t]
\centering
\includegraphics[
    width=\linewidth
]{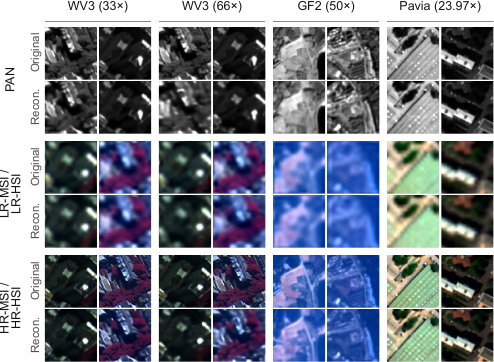}

\caption{\textbf{TOBT multi-modality reconstruction results on WV3, GF2, and Pavia.} Upper rows: original test crops; lower rows: independent TOBT reconstructions. Ratios are overall TOP compression ratios (Eq.~\eqref{compression ratio}).}
\label{fig:exp1_recon_result}
\end{figure}

\subsubsection{Evaluation Metrics}
On scenes with GT, we report PSNR, SSIM~\citep{Wang2004SSIM}, SAM (in degrees)~\citep{Kruse1993SAM}, and ERGAS~\citep{Vivone2015Critical}. We also report the no-reference HQNR index and its spectral and spatial distortion components $D_{\lambda}$ and $D_{s}$~\citep{Aiazzi2014HQNR}. In the reduced-resolution tables, they compare each fused image with the LR-MSI/LR-HSI and PAN of the same scene and thus measure consistency with the source observations rather than full-resolution quality. For the lossy baselines, $D_{\lambda}$ and $D_{s}$ are computed against the original lossless observations rather than their decompressed counterparts. Otherwise, a codec could distort both the fusion input and the metric anchor in a similar direction and receive an artificially favorable consistency score. Retaining the common lossless anchors makes HQNR terms comparable across codecs and TOP, while PSNR, SSIM, SAM, and ERGAS independently measure agreement with GT. In the result tables, the best and second-best results are shown in boldface and underlined, respectively.

\subsubsection{Implementation Details}
\begin{figure*}[!t]
\centering
\includegraphics[width=0.94\textwidth]{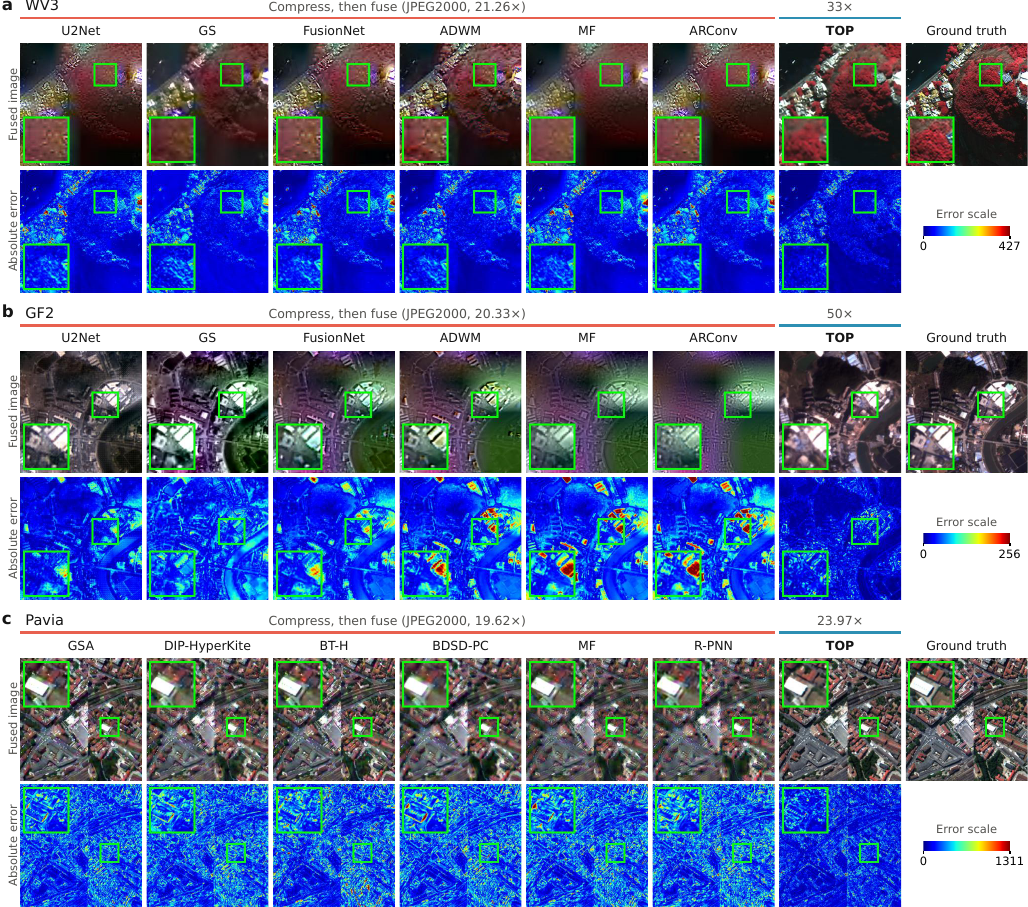}
\caption{\textbf{Qualitative comparison under the JPEG2000 transmission-oriented protocol on (a) WV3, (b) GF2, and (c) Pavia.} TOP is compared with the six baselines of highest PSNR (averaged over WV3 and GF2 for the multispectral data). Rows show the predictions with GT and the error maps (dark blue: lower error). All baselines are compared in Sec.~\suppref{sec:supp_jpeg2000_qualitative} of the Supplementary Material.}
\label{fig:qualitative_comparison}
\end{figure*}

\begin{table*}[!t]
\caption{WV3 Reduced-Resolution Results Under JPEG2000 (Mean $\pm$ Std.)}
\label{tab:main_wv3_jpeg2000}
\centering
\scriptsize
\setlength{\tabcolsep}{\widetabcolsep}
\renewcommand{\arraystretch}{1.0}
\begin{tabular*}{\textwidth}{@{\extracolsep{\fill}}lcccccccc@{}}
\toprule
Method & Ratio $\uparrow$ & PSNR $\pm$ Std. $\uparrow$ & SSIM $\pm$ Std. $\uparrow$ & ERGAS $\pm$ Std. $\downarrow$ & HQNR $\pm$ Std. $\uparrow$ & $D_\lambda$ $\pm$ Std. $\downarrow$ & $D_s$ $\pm$ Std. $\downarrow$ & SAM (deg) $\pm$ Std. $\downarrow$ \\
\midrule
BDSD+JPEG2000~\citep{Garzelli2008BDSD} & \multirow{10}{*}{21.26$\times$} & $25.44\pm3.85$ & $0.55\pm0.16$ & $10.85\pm2.31$ & $0.17\pm0.06$ & $0.68\pm0.06$ & $0.48\pm0.12$ & $11.20\pm2.44$ \\
FusionNet+JPEG2000~\citep{Deng2021FusionNet} & & $26.50\pm3.69$ & $0.73\pm0.09$ & $9.39\pm2.31$ & $0.36\pm0.08$ & $0.54\pm0.08$ & $0.23\pm0.09$ & $10.51\pm2.34$ \\
GLP+JPEG2000~\citep{Aiazzi2002GLP} & & $26.53\pm3.84$ & $0.68\pm0.11$ & $9.56\pm2.04$ & $0.27\pm0.06$ & $0.63\pm0.05$ & $0.29\pm0.11$ & $10.64\pm2.42$ \\
GS+JPEG2000~\citep{Laben2000GS} & & $27.70\pm3.74$ & $0.71\pm0.11$ & $8.36\pm1.78$ & $0.47\pm0.09$ & $0.43\pm0.08$ & $0.18\pm0.09$ & $10.52\pm2.49$ \\
MF+JPEG2000~\citep{Restaino2016MF} & & $26.59\pm3.88$ & $0.67\pm0.12$ & $9.49\pm2.08$ & $0.27\pm0.08$ & $0.62\pm0.07$ & $0.29\pm0.12$ & $10.44\pm2.34$ \\
U2Net+JPEG2000~\citep{Peng2023U2Net} & & $26.62\pm3.63$ & $0.70\pm0.10$ & $9.34\pm2.05$ & $0.33\pm0.08$ & $0.56\pm0.08$ & $0.26\pm0.09$ & $10.39\pm2.23$ \\
WFANet+JPEG2000~\citep{Huang2025WFANet} & & $25.64\pm3.92$ & $0.64\pm0.13$ & $10.54\pm2.45$ & $0.20\pm0.08$ & $0.68\pm0.07$ & $0.38\pm0.11$ & $10.62\pm2.34$ \\
ZS-PAN+JPEG2000~\citep{Cao2024ZSPan} & & $26.02\pm3.54$ & $0.62\pm0.14$ & $9.92\pm1.70$ & $0.23\pm0.09$ & $0.65\pm0.08$ & $0.37\pm0.14$ & $11.00\pm2.44$ \\
ADWM+JPEG2000~\citep{Huang2025ADWM} & & $26.02\pm5.52$ & $0.71\pm0.14$ & $11.23\pm5.66$ & $0.38\pm0.12$ & $0.53\pm0.10$ & $0.21\pm0.12$ & $15.19\pm7.37$ \\
ARConv+JPEG2000~\citep{Wang2025ARConv} & & $26.70\pm3.84$ & $0.70\pm0.11$ & $9.33\pm1.95$ & $0.29\pm0.08$ & $0.60\pm0.07$ & $0.29\pm0.10$ & $10.43\pm2.30$ \\
\midrule
TOP (ours) & 33$\times$ & $\mathbf{30.32}\pm2.83$ & $\mathbf{0.82}\pm0.04$ & $\mathbf{6.00}\pm1.23$ & $\mathbf{0.76}\pm0.14$ & $\mathbf{0.19}\pm0.11$ & $\mathbf{0.08}\pm0.06$ & $\mathbf{6.00}\pm1.38$ \\
TOP (ours) & \textbf{66$\times$} & $\underline{28.88}\pm3.06$ & $\underline{0.75}\pm0.07$ & $\underline{7.13}\pm1.48$ & $\underline{0.69}\pm0.13$ & $\underline{0.23}\pm0.11$ & $\underline{0.11}\pm0.05$ & $\underline{6.69}\pm1.56$ \\
\bottomrule
\end{tabular*}
\end{table*}

TOP is implemented in PyTorch with bfloat16 mixed precision on an NVIDIA RTX 4090 GPU and optimized with AdamW at an initial learning rate of $2\times10^{-4}$, with weight decay 0.05 for TOBT and 0.01 for the latent fusion module. Architectures and training schedules are listed in Sec.~\suppref{sec:supp_implementation} of the Supplementary Material.

\subsection{Multimodal Reconstruction of TOBT}
\label{sec:exp_tobt_reconstruction}
Table~\ref{tab:tobt_reconstruction} reports validation-set reconstruction of PAN, upsampled LR-MSI/LR-HSI, and HR-MSI/HR-HSI images, and Fig.~\ref{fig:exp1_recon_result} shows examples. These modality-wise results isolate the representation capacity available before latent fusion. On WV3, halving the HR-MSI payload from 0.063 to 0.031 bppb lowers reconstruction PSNR from 30.17 to 28.22 dB, exposing the expected rate--fidelity trade-off. LR-MSI reconstruction nevertheless remains above 37 dB and 0.95 SSIM. TOBT also reaches 33.46 dB on GF2 HR-MSI and 34.85 dB on Pavia HR-HSI, supporting the shared design across sensors and spectral dimensionalities. Fine textures become smoother as the ratio increases, while the main structures and spectral appearance remain stable. These results define the tokenizer's reconstruction ceiling; the end-to-end experiments test how closely the predicted fused codes approach it.

\subsection{Transmission-Oriented Pansharpening}
\label{sec:exp_main_results}
Having established that TOBT preserves the three modalities under a constrained binary capacity, we evaluate whether the complete TOP pipeline can turn that representation into a useful downlink product. The comparison therefore considers fusion fidelity and communication cost jointly: a method is favorable only if it preserves spatial--spectral information while reducing the transmitted payload relative to the codec-based pipelines.

\subsubsection{Multispectral Pansharpening on WV3}
\label{sec:exp_wv3}
Table~\ref{tab:main_wv3_jpeg2000} reports the WV3 reduced-resolution results under JPEG2000. The best baseline reaches 27.70 dB at $21.26\times$, whereas TOP achieves 30.32 dB at $33\times$ and 28.88 dB at $66\times$; relative to the best baseline value of each metric, TOP also reduces ERGAS from 8.36 to 6.00 and SAM from 10.39 to 6.00 degrees. Both TOP configurations outperform all baselines on every metric. Fig.~\ref{fig:qualitative_comparison}(a) shows that TOP preserves sharper local structures and more stable spectral appearance, whereas the codec-based baselines suffer from edge degradation and texture loss. Comparisons with all baselines are given in Sec.~\suppref{sec:supp_jpeg2000_qualitative} of the Supplementary Material.

\subsubsection{Multispectral Pansharpening on GF2}
\label{sec:exp_gf2}
\begin{table}[!t]
\caption{TOBT Reconstruction on the Validation Sets (Mean $\pm$ Std.; bppb: Modality-Specific Payload, Eq.~\eqref{bppb})}
\label{tab:tobt_reconstruction}
\centering
\scriptsize
\begin{tabular}{llccc}
\toprule
Dataset & Modality & bppb $\downarrow$ & PSNR $\pm$ Std. $\uparrow$ & SSIM $\pm$ Std. $\uparrow$ \\
\midrule
\multirow{3}{*}{WV3 (33$\times$)} & HR-MSI & 0.063 & $30.17\pm4.42$ & $0.83\pm0.08$ \\
 & LR-MSI & 0.063 & $39.61\pm3.18$ & $0.97\pm0.01$ \\
 & PAN & 0.500 & $34.88\pm5.01$ & $0.91\pm0.04$ \\
\cmidrule(lr){2-5}
\multirow{3}{*}{WV3 (66$\times$)} & HR-MSI & 0.031 & $28.22\pm4.59$ & $0.74\pm0.12$ \\
 & LR-MSI & 0.031 & $37.14\pm3.73$ & $0.95\pm0.02$ \\
 & PAN & 0.250 & $31.94\pm5.43$ & $0.83\pm0.08$ \\
\midrule
\multirow{3}{*}{GF2} & HR-MSI & 0.063 & $33.46\pm3.75$ & $0.87\pm0.06$ \\
 & LR-MSI & 0.063 & $40.31\pm2.85$ & $0.97\pm0.02$ \\
 & PAN & 0.250 & $33.46\pm3.94$ & $0.87\pm0.05$ \\
\midrule
\multirow{3}{*}{Pavia} & HR-HSI & 0.039 & $34.85\pm3.50$ & $0.92\pm0.02$ \\
 & LR-HSI & 0.039 & $43.90\pm2.57$ & $0.99\pm0.00$ \\
 & PAN & 4.000 & $40.86\pm2.93$ & $0.97\pm0.01$ \\
\bottomrule
\end{tabular}
\end{table}

\begin{table*}[!t]
\caption{GF2 Reduced-Resolution Results Under JPEG2000 (Mean $\pm$ Std.)}
\label{tab:main_gf2_jpeg2000}
\centering
\scriptsize
\setlength{\tabcolsep}{\widetabcolsep}
\renewcommand{\arraystretch}{1.0}
\begin{tabular*}{\textwidth}{@{\extracolsep{\fill}}lcccccccc@{}}
\toprule
Method & Ratio $\uparrow$ & PSNR $\pm$ Std. $\uparrow$ & SSIM $\pm$ Std. $\uparrow$ & ERGAS $\pm$ Std. $\downarrow$ & HQNR $\pm$ Std. $\uparrow$ & $D_\lambda$ $\pm$ Std. $\downarrow$ & $D_s$ $\pm$ Std. $\downarrow$ & SAM (deg) $\pm$ Std. $\downarrow$ \\
\midrule
BDSD+JPEG2000~\citep{Garzelli2008BDSD} & \multirow{10}{*}{20.33$\times$} & $24.33\pm2.19$ & $0.63\pm0.08$ & $5.78\pm0.99$ & $0.09\pm0.02$ & $0.80\pm0.02$ & $0.57\pm0.06$ & $5.32\pm0.76$ \\
FusionNet+JPEG2000~\citep{Deng2021FusionNet} & & $26.73\pm2.16$ & $\underline{0.81}\pm0.03$ & $4.36\pm0.74$ & $0.46\pm0.06$ & $0.50\pm0.06$ & $\underline{0.08}\pm0.03$ & $4.97\pm0.80$ \\
GLP+JPEG2000~\citep{Aiazzi2002GLP} & & $24.56\pm2.13$ & $0.69\pm0.08$ & $5.62\pm0.96$ & $0.12\pm0.03$ & $0.79\pm0.03$ & $0.46\pm0.08$ & $5.28\pm0.77$ \\
GS+JPEG2000~\citep{Laben2000GS} & & $26.99\pm1.77$ & $0.78\pm0.05$ & $4.16\pm0.64$ & $0.53\pm0.07$ & $0.36\pm0.06$ & $0.18\pm0.05$ & $5.36\pm1.14$ \\
MF+JPEG2000~\citep{Restaino2016MF} & & $25.04\pm2.18$ & $0.74\pm0.06$ & $5.32\pm0.93$ & $0.20\pm0.04$ & $0.73\pm0.04$ & $0.29\pm0.08$ & $5.20\pm0.77$ \\
U2Net+JPEG2000~\citep{Peng2023U2Net} & & $\underline{28.53}\pm1.65$ & $0.81\pm0.03$ & $\underline{3.41}\pm0.45$ & $\underline{0.63}\pm0.07$ & $\underline{0.31}\pm0.06$ & $0.09\pm0.04$ & $\underline{4.29}\pm0.60$ \\
WFANet+JPEG2000~\citep{Huang2025WFANet} & & $25.65\pm2.23$ & $0.76\pm0.04$ & $4.88\pm0.85$ & $0.27\pm0.03$ & $0.65\pm0.04$ & $0.23\pm0.05$ & $4.99\pm0.76$ \\
ADWM+JPEG2000~\citep{Huang2025ADWM} & & $25.95\pm2.29$ & $0.79\pm0.04$ & $4.74\pm0.89$ & $0.38\pm0.05$ & $0.56\pm0.05$ & $0.13\pm0.05$ & $5.10\pm0.84$ \\
ARConv+JPEG2000~\citep{Wang2025ARConv} & & $24.82\pm2.24$ & $0.72\pm0.06$ & $5.41\pm0.96$ & $0.14\pm0.02$ & $0.77\pm0.02$ & $0.37\pm0.06$ & $5.10\pm0.75$ \\
ZS-PAN+JPEG2000~\citep{Cao2024ZSPan} & & $24.52\pm2.05$ & $0.67\pm0.08$ & $5.63\pm0.90$ & $0.12\pm0.03$ & $0.78\pm0.03$ & $0.47\pm0.08$ & $5.34\pm0.77$ \\
\midrule
TOP (ours) & \textbf{50$\times$} & $\mathbf{31.37}\pm1.90$ & $\mathbf{0.83}\pm0.05$ & $\mathbf{2.63}\pm0.50$ & $\mathbf{0.83}\pm0.03$ & $\mathbf{0.11}\pm0.03$ & $\mathbf{0.07}\pm0.02$ & $\mathbf{2.46}\pm0.42$ \\
\bottomrule
\end{tabular*}
\end{table*}
On GF2 (Tab.~\ref{tab:main_gf2_jpeg2000}), TOP achieves 31.37 dB PSNR, 0.83 SSIM, 2.63 ERGAS, and 2.46 degrees SAM at $50\times$. The strongest baseline, U2Net, reaches 28.53 dB and 4.29 degrees at $20.33\times$, so TOP improves PSNR by 2.84 dB at about 2.5 times its compression ratio; the zero-shot ZS-PAN reaches 24.52 dB. Fig.~\ref{fig:qualitative_comparison}(b) shows that TOP avoids the codec artifacts that affect the baselines.

Tables~\ref{tab:main_wv3_hevc} and~\ref{tab:main_gf2_hevc} report the corresponding reduced-resolution comparisons under HEVC. The TOP rows are identical to those in the JPEG2000 tables because neither codec is part of TOP: HEVC and JPEG2000 are applied only to the source observations transmitted by the pixel-domain baselines, whereas TOP always bit-packs and transmits the same fused binary latent. Repeating the TOP operating point in each table provides a direct comparison against the baseline pipeline induced by that codec. Although the HEVC baselines operate at only $6.69\times$ on WV3 and $8.94\times$ on GF2, TOP exceeds their strongest PSNR values (27.22 and 29.13 dB) by 3.10 and 2.24 dB at $33\times$ and $50\times$, respectively, and leads on every metric. The persistence of these gains across two sensors and two codecs indicates that the advantage comes from changing the transmitted representation rather than adapting to the artifacts of one particular codec. The corresponding visual comparisons are provided in Sec.~\suppref{app:hevc_qualitative}, and full-resolution results, where TOP attains the highest HQNR and lowest $D_\lambda$ under both codecs, are given in Sec.~\suppref{sec:supp_full_resolution_msi} of the Supplementary Material.

\begin{table*}[!t]
\caption{WV3 Reduced-Resolution Results Under HEVC (Mean $\pm$ Std.)}
\label{tab:main_wv3_hevc}
\centering
\scriptsize
\setlength{\tabcolsep}{\widetabcolsep}
\renewcommand{\arraystretch}{0.90}
\begin{tabular*}{\textwidth}{@{\extracolsep{\fill}}lcccccccc@{}}
\toprule
Method & Ratio $\uparrow$ & PSNR $\pm$ Std. $\uparrow$ & SSIM $\pm$ Std. $\uparrow$ & ERGAS $\pm$ Std. $\downarrow$ & HQNR $\pm$ Std. $\uparrow$ & $D_\lambda$ $\pm$ Std. $\downarrow$ & $D_s$ $\pm$ Std. $\downarrow$ & SAM (deg) $\pm$ Std. $\downarrow$ \\
\midrule
BDSD+HEVC & \multirow{10}{*}{6.69$\times$} & $26.53\pm3.65$ & $0.60\pm0.14$ & $9.56\pm2.05$ & $0.31\pm0.08$ & $0.50\pm0.10$ & $0.39\pm0.08$ & $10.72\pm2.52$ \\
FusionNet+HEVC & & $26.59\pm3.55$ & $0.62\pm0.12$ & $9.31\pm2.07$ & $0.34\pm0.09$ & $0.49\pm0.11$ & $0.34\pm0.07$ & $9.71\pm2.36$ \\
GLP+HEVC & & $26.79\pm3.58$ & $0.62\pm0.13$ & $9.24\pm1.97$ & $0.31\pm0.07$ & $0.51\pm0.09$ & $0.37\pm0.07$ & $9.52\pm2.30$ \\
GS+HEVC & & $27.22\pm3.53$ & $0.64\pm0.12$ & $8.82\pm1.91$ & $0.38\pm0.10$ & $0.46\pm0.11$ & $0.31\pm0.06$ & $9.79\pm2.36$ \\
MF+HEVC & & $26.92\pm3.58$ & $0.62\pm0.13$ & $9.13\pm2.00$ & $0.32\pm0.09$ & $0.51\pm0.10$ & $0.35\pm0.07$ & $9.35\pm2.26$ \\
U2Net+HEVC & & $26.65\pm3.54$ & $0.61\pm0.13$ & $9.36\pm2.05$ & $0.33\pm0.09$ & $0.49\pm0.11$ & $0.36\pm0.07$ & $9.48\pm2.29$ \\
WFANet+HEVC & & $26.24\pm3.70$ & $0.59\pm0.15$ & $9.82\pm2.17$ & $0.27\pm0.07$ & $0.55\pm0.09$ & $0.41\pm0.08$ & $9.65\pm2.35$ \\
ZS-PAN+HEVC & & $26.57\pm3.44$ & $0.60\pm0.14$ & $9.40\pm1.85$ & $0.28\pm0.09$ & $0.54\pm0.11$ & $0.40\pm0.08$ & $9.60\pm2.32$ \\
ADWM+HEVC & & $26.83\pm3.70$ & $0.63\pm0.13$ & $9.18\pm2.19$ & $0.37\pm0.09$ & $0.46\pm0.11$ & $0.32\pm0.07$ & $10.13\pm3.12$ \\
ARConv+HEVC & & $26.64\pm3.59$ & $0.61\pm0.14$ & $9.39\pm2.00$ & $0.30\pm0.08$ & $0.52\pm0.10$ & $0.38\pm0.08$ & $9.49\pm2.31$ \\
\midrule
TOP (ours) & 33$\times$ & $\mathbf{30.32}\pm2.83$ & $\mathbf{0.82}\pm0.04$ & $\mathbf{6.00}\pm1.23$ & $\mathbf{0.76}\pm0.14$ & $\mathbf{0.19}\pm0.11$ & $\mathbf{0.08}\pm0.06$ & $\mathbf{6.00}\pm1.38$ \\
TOP (ours) & \textbf{66$\times$} & $\underline{28.88}\pm3.06$ & $\underline{0.75}\pm0.07$ & $\underline{7.13}\pm1.48$ & $\underline{0.69}\pm0.13$ & $\underline{0.23}\pm0.11$ & $\underline{0.11}\pm0.05$ & $\underline{6.69}\pm1.56$ \\
\bottomrule
\end{tabular*}
\end{table*}

\begin{table*}[!t]
\caption{GF2 Reduced-Resolution Results Under HEVC (Mean $\pm$ Std.)}
\label{tab:main_gf2_hevc}
\centering
\scriptsize
\setlength{\tabcolsep}{\widetabcolsep}
\renewcommand{\arraystretch}{0.90}
\begin{tabular*}{\textwidth}{@{\extracolsep{\fill}}lcccccccc@{}}
\toprule
Method & Ratio $\uparrow$ & PSNR $\pm$ Std. $\uparrow$ & SSIM $\pm$ Std. $\uparrow$ & ERGAS $\pm$ Std. $\downarrow$ & HQNR $\pm$ Std. $\uparrow$ & $D_\lambda$ $\pm$ Std. $\downarrow$ & $D_s$ $\pm$ Std. $\downarrow$ & SAM (deg) $\pm$ Std. $\downarrow$ \\
\midrule
BDSD+HEVC & \multirow{10}{*}{8.94$\times$} & $28.03\pm1.91$ & $0.70\pm0.07$ & $3.77\pm0.65$ & $0.43\pm0.06$ & $0.40\pm0.05$ & $0.29\pm0.06$ & $3.95\pm0.59$ \\
FusionNet+HEVC & & $28.68\pm1.84$ & $\underline{0.73}\pm0.06$ & $3.47\pm0.56$ & $0.50\pm0.06$ & $0.37\pm0.05$ & $0.21\pm0.04$ & $3.85\pm0.57$ \\
GLP+HEVC & & $27.78\pm1.86$ & $0.69\pm0.07$ & $3.88\pm0.65$ & $0.37\pm0.06$ & $0.46\pm0.05$ & $0.32\pm0.06$ & $3.91\pm0.57$ \\
GS+HEVC & & $28.89\pm1.71$ & $0.73\pm0.06$ & $3.36\pm0.54$ & $\underline{0.61}\pm0.05$ & $0.31\pm0.04$ & $\underline{0.11}\pm0.05$ & $3.81\pm0.60$ \\
MF+HEVC & & $28.16\pm1.89$ & $0.71\pm0.07$ & $3.72\pm0.62$ & $0.43\pm0.06$ & $0.42\pm0.05$ & $0.26\pm0.06$ & $3.87\pm0.57$ \\
U2Net+HEVC & & $\underline{29.13}\pm1.72$ & $0.72\pm0.06$ & $\underline{3.25}\pm0.52$ & $0.61\pm0.05$ & $\underline{0.29}\pm0.04$ & $0.14\pm0.04$ & $\underline{3.55}\pm0.49$ \\
WFANet+HEVC & & $28.05\pm1.91$ & $0.69\pm0.07$ & $3.72\pm0.61$ & $0.41\pm0.05$ & $0.42\pm0.05$ & $0.29\pm0.05$ & $3.92\pm0.56$ \\
ADWM+HEVC & & $28.52\pm1.87$ & $0.72\pm0.06$ & $3.55\pm0.58$ & $0.47\pm0.05$ & $0.38\pm0.05$ & $0.24\pm0.04$ & $3.86\pm0.57$ \\
ARConv+HEVC & & $27.50\pm1.98$ & $0.67\pm0.08$ & $3.98\pm0.68$ & $0.34\pm0.06$ & $0.47\pm0.05$ & $0.36\pm0.06$ & $3.91\pm0.57$ \\
ZS-PAN+HEVC & & $27.72\pm1.85$ & $0.69\pm0.07$ & $3.88\pm0.64$ & $0.37\pm0.05$ & $0.45\pm0.05$ & $0.33\pm0.05$ & $3.93\pm0.57$ \\
\midrule
TOP (ours) & \textbf{50$\times$} & $\mathbf{31.37}\pm1.90$ & $\mathbf{0.83}\pm0.05$ & $\mathbf{2.63}\pm0.50$ & $\mathbf{0.83}\pm0.03$ & $\mathbf{0.11}\pm0.03$ & $\mathbf{0.07}\pm0.02$ & $\mathbf{2.46}\pm0.42$ \\
\bottomrule
\end{tabular*}
\end{table*}

\subsubsection{Hyperspectral Pansharpening on Pavia}

\begin{table*}[!t]
\caption{Pavia Results Under JPEG2000 (Mean over Two \texttt{Full} Test Scenes; Std. Omitted)}
\label{tab:main_hs}
\centering
\scriptsize
\setlength{\tabcolsep}{\widetabcolsep}
\renewcommand{\arraystretch}{1.0}
\begin{tabular*}{\textwidth}{@{\extracolsep{\fill}}lcccccccc@{}}
\toprule
Method & Ratio $\uparrow$ & \makecell[c]{PSNR $\uparrow$} & \makecell[c]{SSIM $\uparrow$} & \makecell[c]{ERGAS $\downarrow$} & \makecell[c]{HQNR $\uparrow$} & \makecell[c]{$D_\lambda$ $\downarrow$} & \makecell[c]{$D_s$ $\downarrow$} & \makecell[c]{SAM (deg) $\downarrow$} \\
\midrule
AWLP+JPEG2000~\citep{Otazu2005AWLP}
& \multirow{12}{*}{19.62$\times$}
& $26.15$
& $0.65$
& $8.94$
& $0.76$
& $0.16$
& $0.09$
& $10.91$ \\

BDSD-PC+JPEG2000~\citep{Vivone2019BDSDPC}
& & $27.02$
& $0.65$
& $8.15$
& $\underline{0.83}$
& $0.15$
& $\underline{0.03}$
& $9.69$ \\

BT-H+JPEG2000~\citep{Lolli2017HazeCorrection}
& & $27.25$
& $\underline{0.72}$
& $8.60$
& $0.66$
& $0.30$
& $0.06$
& $10.16$ \\

GSA+JPEG2000~\citep{Aiazzi2007GSA}
& & $\underline{28.09}$
& $0.69$
& $\underline{7.54}$
& $0.78$
& $0.16$
& $0.07$
& $10.20$ \\

MF+JPEG2000~\citep{Restaino2016MF}
& & $26.72$
& $0.67$
& $8.39$
& $0.78$
& $0.16$
& $0.07$
& $10.00$ \\

PRACS+JPEG2000~\citep{Choi2011PRACS}
& & $26.58$
& $0.62$
& $8.59$
& $0.81$
& $0.15$
& $0.05$
& $9.73$ \\

uTDSP+JPEG2000~\citep{Jiang2025uTDSP}
& & $25.26$
& $0.48$
& $10.28$
& $0.69$
& $0.15$
& $0.19$
& $12.40$ \\

DIP-HyperKite+JPEG2000~\citep{Bandara2022DIPHyperKite}
& & $27.40$
& $0.69$
& $7.80$
& $0.82$
& $0.15$
& $0.03$
& $\underline{9.68}$ \\

HyperTransformer+JPEG2000~\citep{Bandara2022HyperTransformer}
& & $19.19$
& $0.21$
& $20.50$
& $0.44$
& $0.25$
& $0.42$
& $34.43$ \\

R-PNN+JPEG2000~\citep{Guarino2024RPNN}
& & $26.62$
& $0.64$
& $8.46$
& $0.79$
& $0.15$
& $0.06$
& $10.34$ \\

DM-ZS+JPEG2000~\citep{Xiao2025DMZS}
& & $26.18$
& $0.67$
& $9.67$
& $0.75$
& $0.20$
& $0.06$
& $11.70$ \\

$\rho$-PNN+JPEG2000~\citep{Guarino2025RhoPNN}
& & $25.14$
& $0.58$
& $10.00$
& $0.75$
& $\underline{0.14}$
& $0.13$
& $11.31$ \\

\midrule
TOP (ours)
& \textbf{23.97$\times$}
& $\mathbf{29.72}$
& $\mathbf{0.83}$
& $\mathbf{5.91}$
& $\mathbf{0.85}$
& $\mathbf{0.14}$
& $\mathbf{0.02}$
& $\mathbf{6.39}$ \\
\bottomrule
\end{tabular*}
\end{table*}

Table~\ref{tab:main_hs} reports the Pavia results under JPEG2000. The baselines operate at $19.62\times$, with the best PSNR and HQNR reaching 28.09 dB and 0.83. At the higher ratio of $23.97\times$, TOP achieves 29.72 dB PSNR, 0.83 SSIM, 5.91 ERGAS, 0.85 HQNR, 6.39 degrees SAM, $D_{\lambda}=0.14$, and $D_s=0.02$, the best result on every metric. Fig.~\ref{fig:qualitative_comparison}(c) shows that TOP also preserves hyperspectral spatial structures under a constrained downlink. The HEVC results, where TOP again outperforms all baselines, are given in Sec.~\suppref{app:hevc_qualitative} of the Supplementary Material. Across WV3, GF2, and Pavia, the same pattern holds for 8-, 4-, and 102-band data: TOP improves both GT-based fidelity and consistency with the source observations while using a higher compression ratio. The simultaneous gains in PSNR, SAM, $D_{\lambda}$, and $D_s$ are important because a higher PSNR alone could mask band-dependent spectral distortion. The following profile analysis examines this balance directly at the band level.

\subsubsection{Spectral Profile Analysis}

\begin{figure*}[!t]
\centering
\includegraphics[width=\textwidth]{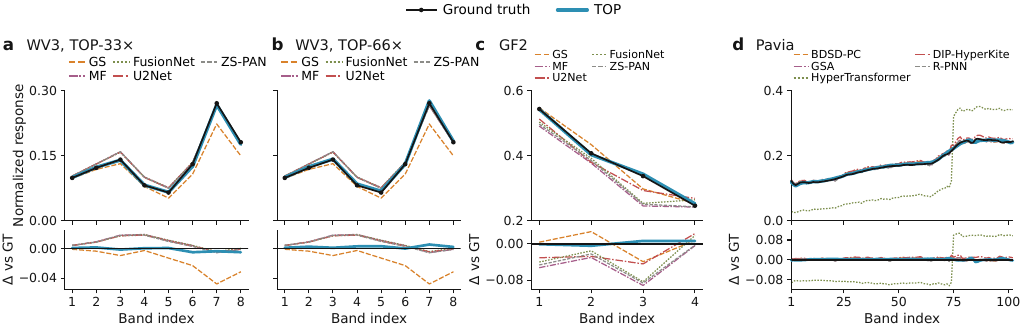}
\caption{\textbf{Spectral profile comparison on WV3, GF2, and Pavia.} (a, b) WV3 with TOP at 33$\times$ and 66$\times$, (c) GF2 at 50$\times$, and (d) Pavia at 23.97$\times$. Upper plots: normalized band-wise mean over a fixed ROI; lower strips: deviation from GT; best-ranked baseline of each method family.}
\label{fig:Overall_Spectral_Response}
\end{figure*}

Image-level metrics summarize the final reconstruction but do not show how compression and fusion redistribute errors across spectral bands. Fig.~\ref{fig:Overall_Spectral_Response} therefore compares the normalized band-wise mean over a fixed ROI with the GT profile, with all baselines paired with JPEG2000. In all four settings, the TOP curve follows the GT profile most closely and its residual strip remains nearer to zero, whereas the codec-based baselines show larger band-dependent amplitude errors. This trend persists for WV3 at both $33\times$ and $66\times$ and extends to GF2 and the 102-band Pavia data. The spectral-profile evidence thus complements SAM and $D_\lambda$: the quantitative gains are not produced by spatial sharpening at the expense of spectral consistency.

\section[Ablation Studies]{Ablation Studies}\label{sec:exp_ablation}
\begin{figure*}[!t]
\centering
\includegraphics[width=\textwidth]{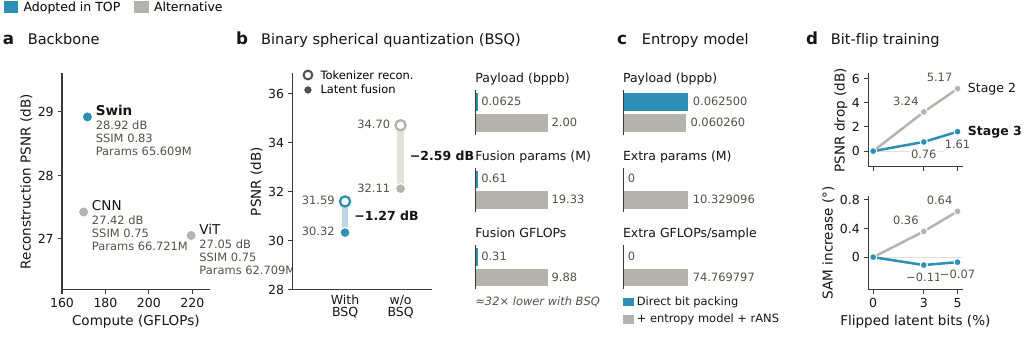}
\caption{\textbf{Ablation studies of the key design choices of TOP.} WV3 reduced-resolution data; blue: design adopted in TOP, gray: alternative. (a) Backbone (HR-MSI reconstruction). (b) BSQ: reconstruction--fusion PSNR gap (left), payload and fusion-module complexity (right). (c) Additional entropy model with rANS. (d) Degradation under random bit flips after stages 2 and 3. Bars in (b) and (c) are scaled per pair. Values are listed in Sec.~\suppref{sec:supp_ablation} of the Supplementary Material.}
\label{fig:ablation_summary}
\end{figure*}

We ablate four design choices of TOP on the WV3 reduced-resolution data: the tokenizer backbone, BSQ, an additional entropy model, and stage-3 bit-flip training. These components address different system constraints, so they cannot be judged by reconstruction accuracy alone: the backbone controls representation quality and computation, BSQ affects payload and latent predictability, entropy coding trades bitrate for sender-side overhead, and stage~3 controls sensitivity to erroneous bits. Fig.~\ref{fig:ablation_summary} summarizes these accuracy--payload--complexity trade-offs, and the numerical values are listed in panels (a)--(e) of Tab.~\suppref{tab:ablation_summary} in the Supplementary Material.

\subsection{Architecture Choice}\label{sec:exp_ablation_arch}
We compare CNN, Vision Transformer (ViT)~\citep{Dosovitskiy2021ViT}, and Swin~\citep{Liu2021Swin} backbones with 66.721M, 62.709M, and 65.609M parameters, respectively, keeping the latent configuration, quantization, and training fixed; this Swin variant is matched to the parameter budget and differs from the dataset-specific tokenizer configurations listed in Tab.~\suppref{tab:implementation_architecture} of the Supplementary Material. As shown in Fig.~\ref{fig:ablation_summary}(a), Swin achieves the best HR-MSI reconstruction at a computational cost close to that of the CNN and lower than that of ViT, whereas ViT adds computation without a corresponding gain. Swin is therefore adopted for TOBE and TOBD.

\subsection{Effectiveness of BSQ}\label{sec:exp_ablation_bsq}
BSQ reduces the payload to one bit per latent entry (Fig.~\ref{fig:ablation_summary}(b)). To test whether it also makes the target latent easier to predict, we compare the fusion PSNR with the PSNR of directly encoding and decoding the GT image with the same tokenizer; their gap indicates how closely latent fusion approaches the reconstruction capability of the latent space. Both PSNRs are computed on the 20 WV3 reduced-resolution test scenes, so the fusion PSNR with BSQ is the 30.32 dB of TOP at $33\times$ in Tab.~\ref{tab:main_wv3_jpeg2000}, whereas Tab.~\ref{tab:tobt_reconstruction} reports reconstruction on the validation set. Removing BSQ improves reconstruction, but even with a deliberately enlarged and fully trained fusion module, the continuous-latent variant leaves a 2.59 dB gap between reconstruction and fusion. BSQ reduces this gap to 1.27 dB while shrinking the fusion module from 19.33M to 0.61M parameters and from 9.88 to 0.31 giga floating-point operations (GFLOPs). BSQ thus trades part of the reconstruction capacity for a compact payload and a more predictable fusion target.

\subsection{Trade-Off of an Additional Entropy Model}\label{sec:exp_ablation_entropy}
A neural entropy model, which drives rANS~\citep{Duda2013ANS} or arithmetic coding with estimated symbol probabilities~\citep{Balle2018ScaleHyperprior}, is most effective when the latent retains considerable statistical redundancy, whereas TOBT already compresses the HR information into a tight binary bottleneck. We therefore code the fused binary latent that TOP downlinks for WV3, which uses 0.0625 bppb with direct bit packing, with a recent 10.329096M-parameter dictionary-based entropy model~\citep{Lu2025DictionaryEntropy} and rANS. As shown in Fig.~\ref{fig:ablation_summary}(c), the payload decreases only from 0.062500 to 0.060260 bppb (about 3.58\%), at the cost of 74.769797 additional GFLOPs/sample for probability prediction alone, plus GPU synchronization, data transfer, CPU rANS coding, and packet assembly on the sender side. This overhead is disproportionate to the gain, which supports the entropy-coding-free design of TOP.

\subsection{Stage-3 Training Strategy on Robustness}
\label{sec:exp_ablation_robustness}
The fused code is both the output of onboard fusion and the downlink payload, so a wrong bit may come from imperfect latent prediction or from transmission errors. We randomly flip 3\% or 5\% of the signs of the fused code on the 20 WV3 reduced-resolution test scenes and decode it with TOBD after stage 2 or stage 3, keeping the encoder, latent fusion module, and unperturbed code unchanged. As shown in Fig.~\ref{fig:ablation_summary}(d), stage-3 training reduces the PSNR drop from 3.24 to 0.76 dB at 3\% flips and from 5.17 to 1.61 dB at 5\% flips. Under the same flips, stage 2 increases SAM by 0.36 and 0.64 degrees, whereas after stage 3 SAM changes by only $-0.11$ and $-0.07$ degrees. The small negative values can be interpreted as negligible sample-level variation rather than a benefit from corruption; the relevant result is that spectral fidelity no longer deteriorates systematically. Bit-flip training therefore makes TOBD markedly less sensitive to bit errors.

\section{Discussion}\label{sec:discussion}

\subsection{Mechanism of Latent-Space Fusion}
TOP works because one binary latent space serves reconstruction, fusion, and transmission. In stage~1, the shared tokenizer must reconstruct PAN, LR-MSI/LR-HSI, and HR-MSI/HR-HSI images under the same binary capacity, so it keeps the structures needed for both spatial detail and spectral recovery and discards pixel-level redundancy that is costly to transmit. In this space, the latent fusion module only has to predict a binary state at each aligned latent location, instead of synthesizing every output pixel and band.

The BSQ ablation shows the cost and the benefit of this design (Sec.~\ref{sec:exp_ablation_bsq}). Without BSQ, the tokenizer reconstructs better on its own, but the fusion target is continuous. It then needs a much larger fusion module and still leaves a larger reconstruction--fusion gap. BSQ gives up part of the standalone reconstruction capacity and obtains a target that is compact and easy to predict. The training stages build on this. Stage~2 fine-tunes TOBD on the HR code, and stage~3 adds bit-flip augmentation so that TOBD tolerates wrong bits from imperfect latent prediction or from the channel. With TOBT frozen, latent-fusion training then makes the onboard prediction match the HR code. The fused code is therefore already the payload for the channel, not an intermediate image that must be compressed again.

\subsection{Improvements Over Existing Pipelines}
Ground-side fusion after compression and onboard pixel-domain fusion before compression are both limited by an objective mismatch. In the first, the codec discards information before fusion, and the fusion model is usually trained on clean inputs, so a better fusion model cannot recover what the downlink has already removed. In the second, the fusion model produces a full-resolution image without knowing how a generic codec will quantize it, and the codec minimizes image distortion without knowing which structures matter for pansharpening. Either way, two stages with different objectives are concatenated.

TOP removes this mismatch by making the fused latent code the payload. Tokenization sets what can be represented within the bit budget, onboard fusion predicts that representation, and the ground decoder reconstructs the image from the same code that was optimized during training. This agrees with the results against compress-then-fuse baselines in the main text and against onboard fusion followed by JPEG2000 or HEVC in Sec.~\suppref{sec:supp_onboard_fusion} of the Supplementary Material. The gain therefore comes from coordinating representation, fusion, and transmission, not from one fusion network or one codec alone.

\subsection{Deployment Flexibility}
The binary payload keeps onboard processing simple. It ends with sign selection and bit packing, and it needs neither full-resolution reconstruction nor a learned entropy model (Sec.~\ref{sec:exp_ablation_entropy}). The fixed payload size is known before transmission, which helps buffer allocation and link scheduling, and decoding stays on the ground.

The shared encoder supports two transmission policies. By default, TOP sends the fused latent. If downstream tasks need the source observations, the satellite can instead send the PAN and spectral latents, and the ground station reconstructs them and applies an existing fusion method. Table~\ref{tab:tobt_compressor_only_ground_fusion} shows that this policy remains viable: on TOBT-decoded inputs, FusionNet reaches 29.95 dB on GF2 at $25.0\times$ and 29.32 dB on WV3 at $16.5\times$. These ratios are about half those of TOP because two equal-sized source latents are sent. TOP sends only the fused latent, at $50\times$ and $33\times$, while improving PSNR by 1.42 dB on GF2 and 1.00 dB on WV3. The choice is therefore explicit: transmit the source latents for reuse, or transmit the fused latent for the best rate--quality trade-off.

\begin{table}[!t]
\caption{Ground-Side Fusion After TOBT-Only Transmission (Mean $\pm$ Std.)}
\label{tab:tobt_compressor_only_ground_fusion}
\centering
\resizebox{\linewidth}{!}{
\begin{tabular}{@{}llcccc@{}}
\toprule
Dataset
& Method
& Ratio $\uparrow$
& PSNR $\uparrow$
& SAM (deg) $\downarrow$
& HQNR $\uparrow$ \\
\midrule

\multirow{3}{*}{GF2}
& FusionNet~\citep{Deng2021FusionNet}
& 25.0$\times$
& $29.95\pm1.63$
& $3.33\pm0.48$
& $0.72\pm0.04$ \\

& WFANet~\citep{Huang2025WFANet}
& 25.0$\times$
& $29.19\pm1.77$
& $3.48\pm0.47$
& $0.66\pm0.03$ \\

& TOP (ours)
& \textbf{50$\times$}
& $\mathbf{31.37}\pm1.90$
& $\mathbf{2.46}\pm0.42$
& $\mathbf{0.83}\pm0.03$ \\

\midrule

\multirow{3}{*}{WV3}
& FusionNet~\citep{Deng2021FusionNet}
& 16.5$\times$
& $29.32\pm2.62$
& $7.93\pm1.86$
& $0.65\pm0.15$ \\

& WFANet~\citep{Huang2025WFANet}
& 16.5$\times$
& $27.65\pm3.24$
& $8.24\pm1.94$
& $0.49\pm0.11$ \\

& TOP (ours)
& \textbf{33$\times$}
& $\mathbf{30.32}\pm2.83$
& $\mathbf{6.00}\pm1.38$
& $\mathbf{0.76}\pm0.14$ \\

\bottomrule
\end{tabular}
}
\end{table}

TOP also has two practical limitations: (1) the tokenizer is trained per sensor, and (2) training takes several stages. Sec.~\suppref{sec:supp_limitations} of the Supplementary Material discusses them.

\section{Conclusion}\label{sec:conclusion}

This work presents TOP, which treats pansharpening for satellite downlink as a joint design of compact latent coding, onboard fusion, and transmission. TOBE with a BSQ bottleneck produces a binary latent code that can be reconstructed, predicted, and bit-packed directly. TOBD decodes the fused code on the ground and can tolerate occasional bit errors. On WV3, GF2, and Pavia datasets, TOP produces higher fusion quality than pansharpening methods paired with common codecs, at compression ratios 1.2 to 9.9 times those of these baselines. For example, on WV3 it reaches 30.32 dB PSNR at $33\times$, whereas the best JPEG2000 baseline reaches 27.70 dB at $21.26\times$. These findings suggest that future remote-sensing
systems could benefit from co-designing task processing
and transmission, rather than treating downlink transmission
as a separate stage. Future work will simplify the training procedure and improve generalization across sensors.

\bibliographystyle{IEEEtran}
\bibliography{refs}

\clearpage
\setcounter{section}{0}
\setcounter{subsection}{0}
\setcounter{equation}{19}
\setcounter{figure}{8}
\setcounter{table}{8}
\renewcommand{\thesection}{S\arabic{section}}
\renewcommand{\thesubsection}{\thesection.\arabic{subsection}}
\renewcommand{\theHsection}{supp.\arabic{section}}
\renewcommand{\theHsubsection}{\theHsection.\arabic{subsection}}
\renewcommand{\theHsubsubsection}{\theHsubsection.\arabic{subsubsection}}
\setlength{\widetabcolsep}{3pt}
\twocolumn[{\centering
{\LARGE Supplementary Material for\\[0.4em]
TOP: Transmission-Oriented Remote Sensing Pansharpening\par}
\vspace{1em}
{\large Zi-Han Cao, Jing-Ze Liang, Shao-Jie Wu, Jin-Cheng Di, Liang-Jian Deng, and Gemine Vivone\par}
\vspace{1.5em}}]

\noindent This supplement details BSQ and the latent-fusion loss (Secs.~\ref{supp:sec:supp_bsq}--\ref{supp:app:bce_logits}), implementation and evaluation protocols (Secs.~\ref{supp:sec:supp_implementation}--\ref{supp:sec:supp_datasets}), and additional JPEG2000, HEVC, full-resolution, ablation, and onboard-fusion results (Secs.~\ref{supp:sec:supp_jpeg2000_qualitative}--\ref{supp:sec:supp_onboard_fusion}). Sec.~\ref{supp:sec:supp_limitations} discusses limitations.

\noindent In all tables, bold and underlined entries are the best and second best; repeated-sample results are mean $\pm$ standard deviation. Baseline references appear in the main manuscript. In reduced-resolution figures, prediction and error-map rows alternate; darker blue indicates lower error, and green boxes mark enlarged regions.

\section{Details of Binary Spherical Quantization}
\label{supp:sec:supp_bsq}
\begingroup
\setlength{\abovedisplayskip}{3pt}
\setlength{\belowdisplayskip}{3pt}
\setlength{\jot}{1pt}
Following BSQ~\citep{Zhao2025BSQ}, inference and downlink use only the hard normalized tensor $\mathbf{B}$; the soft assignments below are training auxiliaries.

\paragraph{Hard Spherical Binary Code}
For $\mathbf{f}_{i}\in\mathbb{R}^{\latentdim}$ at latent location $i$,
\begin{equation}
\begin{aligned}
\mathbf{b}_{i}&=\operatorname{sign}(\mathbf{f}_{i})\in\{-1,+1\}^{\latentdim},\\
\mathbf{B}&=\frac{1}{\sqrt{\latentdim}}\operatorname{reshape}([\mathbf{b}_1,\ldots,\mathbf{b}_N])\\
&\in\{-1/\sqrt{\latentdim},+1/\sqrt{\latentdim}\}^{\latentdim\times H/s\times W/s},
\end{aligned}
\end{equation}
where $N=HW/s^2$ and $\operatorname{reshape}$ maps the $\latentdim\times N$ raster-ordered vectors to the latent grid. For each of the three modalities, define
\begin{equation}
\bar{\mathbf{f}}_{i}=\frac{\mathbf{f}_{i}}{\|\mathbf{f}_{i}\|_{2}+\epsilon},
\qquad
\mathbf{q}_{i}=\frac{\mathbf{b}_{i}}{\sqrt{\latentdim}},
\end{equation}
with numerical constant $\epsilon>0$. Positive normalization preserves the sign, and the straight-through estimator (STE) is
\begin{equation}
\mathcal{Q}(\bar{\mathbf{f}}_{i})=\bar{\mathbf{f}}_{i}+\mathrm{sg}(\mathbf{q}_{i}-\bar{\mathbf{f}}_{i}),
\qquad
\frac{\partial\mathcal{Q}}{\partial\bar{\mathbf{f}}_{i}}\overset{\mathrm{STE}}{=}\mathbf{I}_{\latentdim}.
\end{equation}
Thus, the forward value is $\mathbf{q}_i$, while the identity surrogate Jacobian is used only for back-propagation; $\mathrm{sg}$ denotes stop-gradient. Bit packing stores signs, and unpacking restores the two normalized values. The corresponding implicit, unlearned codebook is
\begin{equation}
\mathcal{C}_{\latentdim}=\left\{\frac{\mathbf{c}_{n}}{\sqrt{\latentdim}}\ \middle|\
\begin{array}{l}\mathbf{c}_{n}\in\{-1,+1\}^{\latentdim},\\[-2pt] n=1,\ldots,2^{\latentdim}\end{array}\right\}.
\end{equation}

\paragraph{Geometry and Packed Payload}
The spherical normalization makes binary disagreement directly measurable. For two codewords, with Hamming distance $h(\mathbf b_i,\mathbf b'_i)$,
\begin{equation}
\mathbf q_i^{\top}\mathbf q'_i=1-\frac{2h(\mathbf b_i,\mathbf b'_i)}{\latentdim},
\qquad
\|\mathbf q_i-\mathbf q'_i\|_2^2=\frac{4h(\mathbf b_i,\mathbf b'_i)}{\latentdim}.
\end{equation}
A sign error therefore causes a fixed, bounded latent displacement, and latent prediction errors have the same form as transmission bit errors. The packed tensor contains $N\latentdim=\latentdim HW/s^2$ bits, giving $\mathrm{bppb}=\latentdim/(Cs^2)$ for $C$ target bands. Its payload is fixed by $(\latentdim,s)$ and requires no codebook indices, floating-point magnitudes, or probability tables. The dot-product identity also keeps similarity in the fixed interval $[-1,1]$ across latent dimensions, so changing the rate configuration does not change the scale presented to the decoder.

\paragraph{Soft Assignment for Entropy Regularization}
During training, inverse temperature $\eta>0$ defines the nearest-code distribution
\begin{equation}
\begin{aligned}
p_{i,n}
&=\frac{\exp(-\eta\|\bar{\mathbf{f}}_{i}-\widehat{\mathbf{c}}_{n}\|_2^2)}
{\sum_{j=1}^{2^{\latentdim}}\exp(-\eta\|\bar{\mathbf{f}}_{i}-\widehat{\mathbf{c}}_{j}\|_2^2)}\\
&=\frac{\exp(2\eta\widehat{\mathbf{c}}_{n}^{\top}\bar{\mathbf{f}}_{i})}
{\sum_{j=1}^{2^{\latentdim}}\exp(2\eta\widehat{\mathbf{c}}_{j}^{\top}\bar{\mathbf{f}}_{i})},
\quad \widehat{\mathbf{c}}_{n}=\mathbf{c}_{n}/\sqrt{\latentdim}.
\end{aligned}
\end{equation}
The second form follows because codeword-independent squared-norm terms cancel. The Cartesian binary codebook factorizes by channel:
\begin{equation}
\begin{aligned}
q_{i,d}^{-}&=\sigma\!\left(-\frac{4\eta}{\sqrt{\latentdim}}\bar f_{i,d}\right),
& q_{i,d}^{+}&=1-q_{i,d}^{-},\\[-2pt]
\sigma(z)&=(1+e^{-z})^{-1}.&&
\end{aligned}
\end{equation}
In practice, split the channels into $G=\latentdim/g$ disjoint groups $\mathcal I_h$ of $g$ channels. Let $\bar{\mathbf f}_i^{(h)}=(\bar f_{i,d})_{d\in\mathcal I_h}$, $\mathbf c_{h,n}\in\{-1,+1\}^{g}$ for $n=1,\ldots,2^g$, and $\widehat{\mathbf c}_{h,n}=\mathbf c_{h,n}/\sqrt{\latentdim}$. Then
\begin{equation}
p_{i,h,n}=\frac{\exp(2\eta\widehat{\mathbf c}_{h,n}^{\top}\bar{\mathbf f}_i^{(h)})}
{\sum_{j=1}^{2^g}\exp(2\eta\widehat{\mathbf c}_{h,j}^{\top}\bar{\mathbf f}_i^{(h)})},
\qquad \mathbf p_{i,h}=(p_{i,h,n})_{n=1}^{2^g}.
\end{equation}
This marginal preserves the global $1/\sqrt{\latentdim}$ scaling and reduces explicit assignment from $2^{\latentdim}$ states to $G2^g$ grouped states.

\paragraph{Entropy and Commitment Terms}
For the batch and latent locations $\Omega_{\mathrm e}$,
\begin{equation}
\begin{aligned}
\mathcal H_{\mathrm{sample}}&=\frac{1}{|\Omega_{\mathrm e}|}\sum_{i\in\Omega_{\mathrm e}}\sum_{h=1}^{G}\mathcal H(\mathbf p_{i,h}),\\[-2pt]
\mathcal H(\mathbf p)&=-\sum_n p_n\log(p_n+\epsilon).
\end{aligned}
\end{equation}
while the entropy of average code usage is
\begin{equation}
\bar p_{h,n}=\frac{1}{|\Omega_{\mathrm e}|}\sum_{i\in\Omega_{\mathrm e}}p_{i,h,n},
\qquad \mathcal H_{\mathrm{code}}=\sum_{h=1}^{G}\mathcal H(\bar{\mathbf p}_h),
\end{equation}
where $\bar{\mathbf p}_h=(\bar p_{h,n})_{n=1}^{2^g}$. The first term favors decisive local assignments; the second prevents collapse through broad group-code usage. TOP minimizes
\begin{equation}
\mathcal L_{\mathrm{entropy}}=\frac{\zeta}{\eta}
(\gamma_0\mathcal H_{\mathrm{sample}}-\gamma\mathcal H_{\mathrm{code}})
\end{equation}
and the commitment term
\begin{equation}
\mathcal L_{\mathrm{commit}}=\frac{1}{|\Omega_{\mathrm q}|}\sum_{i\in\Omega_{\mathrm q}}
\|\mathrm{sg}(\mathbf q_i)-\bar{\mathbf f}_i\|_2^2,
\end{equation}
where $\Omega_{\mathrm q}$ is the full batch and latent grid. This moves the continuous feature toward the fixed hard code. Their combination is Eq.~(\mainref{eq:quantization_loss}) of the main manuscript. Tab.~\ref{supp:tab:bsq_parameters} gives all training coefficients; they do not change the deterministic hard code. The effective coefficients of $\mathcal L_{\mathrm{commit}}$, $\mathcal H_{\mathrm{sample}}$, and $\mathcal H_{\mathrm{code}}$ are $(5\!\times\!10^{-3},5\!\times\!10^{-4},-5\!\times\!10^{-4})$ for WV3 $33\times$, $(10^{-2},10^{-3},-10^{-3})$ for WV3 $66\times$, and $(0.1,0.1,-0.1)$ for GF2 and Pavia.

\begin{table}[!t]
\centering
\caption{BSQ training coefficients; training uses hard BSQ with STE, whereas inference uses hard BSQ only.}
\label{supp:tab:bsq_parameters}
\scriptsize
\setlength{\tabcolsep}{2.5pt}
\renewcommand{\arraystretch}{0.88}
\resizebox{\columnwidth}{!}{%
\begin{tabular}{@{}lcccc@{}}
\toprule
Coefficient & WV3 $33\times$ & WV3 $66\times$ & GF2 $50\times$ & Pavia $23.97\times$ \\
\midrule
$\beta,\gamma_0,\gamma,\zeta$ & 0.1 & 0.1 & 1 & 1 \\
$\eta$ & 1 & 1 & 1 & 1 \\
$\lambda_{\mathrm Q}$ & 0.05 & 0.1 & 0.1 & 0.1 \\
\bottomrule
\end{tabular}}
\end{table}

\paragraph{Boundary With Entropy Coding}
The entropy regularizer is used only for optimization; it produces no probability model, side information, or arithmetic-coded stream. Deployment discards $p_{i,n}$, $q_{i,d}^{\pm}$, and $p_{i,h,n}$ and directly bit-packs the signs of $\widehat{\mathbf B}_{\mathrm H}$. A neural entropy model would additionally require probability inference, packet coding, and decoding.
\endgroup

\section{Binary Cross-Entropy Loss of the Latent Fusion Module}
\label{supp:app:bce_logits}
\begingroup
\setlength{\abovedisplayskip}{3pt}
\setlength{\belowdisplayskip}{3pt}
\setlength{\jot}{1pt}
Sec.~\mainref{sec:method_latent_fusion} defines the single logit tensor $\mathbf O$. Its first axis $a\in\{0,1\}$ represents the negative or positive state, and $O_{a,d,u,v}$ is one logit. For each target entry $B_{\mathrm H,d,u,v}$, define
\begin{equation}
y_{d,u,v}=\mathbb I[B_{\mathrm H,d,u,v}>0]\in\{0,1\}
\end{equation}
and lift it to two complementary channels:
\begin{equation}
T_{a,d,u,v}=\mathbb I[a=y_{d,u,v}],\qquad a\in\{0,1\}.
\end{equation}
Thus, $T_{0,d,u,v}=1-y_{d,u,v}$ and $T_{1,d,u,v}=y_{d,u,v}$. The resulting $\mathbf T\in\{0,1\}^{2\times\latentdim\times H/s\times W/s}$ matches $\mathbf O$, so $f_{\mathrm{BCEWithLogits}}(\mathbf O,\mathbf T)$ requires no broadcasting. For scalar logit $o$ and target $t$,
\begin{equation}
\ell(o,t)=\max(o,0)-ot+\log(1+\exp(-|o|)),
\end{equation}
which evaluates the exponential only at a nonpositive argument and avoids an explicit sigmoid. The latent-fusion objective is
\begin{equation}
\begin{aligned}
\mathcal L_{\mathrm{fuse}}&=\frac{1}{2\latentdim N}
\sum_{a=0}^{1}\sum_{d=1}^{\latentdim}\sum_{u=1}^{H/s}\sum_{v=1}^{W/s}
\ell(O_{a,d,u,v},T_{a,d,u,v}),\\[-2pt]
N&=HW/s^2.
\end{aligned}
\end{equation}
This is complementary one-vs-rest BCE, rather than softmax cross-entropy. The frozen values $-1/\sqrt{\latentdim}$ and $+1/\sqrt{\latentdim}$ map to $a=0$ and $a=1$. At inference, $a^*_{d,u,v}=\arg\max_{a\in\{0,1\}}O_{a,d,u,v}$ and $\widehat B_{\mathrm H,d,u,v}=(2a^*_{d,u,v}-1)/\sqrt{\latentdim}$, preserving the training state convention; no sigmoid precedes $f_{\mathrm{BCEWithLogits}}$.
\endgroup

\section{Implementation Details}
\label{supp:sec:supp_implementation}
All configurations use Swin-based encoders/decoders, grouped BSQ with eight channels per group, and a band-adaptive $2\times2$ stride-2 embedding. The kernel layout is output$\times$input$\times$height$\times$width and is interpolated along the input-channel dimension. The factor $s$ applies to both spatial axes, and stage counts include stages without scale changes. Tab.~\ref{supp:tab:implementation_architecture} retains all architecture and configuration-specific training values.

\begin{table}[!t]
\centering
\caption{TOP architecture and configuration-specific training settings.}
\label{supp:tab:implementation_architecture}
\label{supp:tab:implementation_schedule}
\scriptsize
\setlength{\tabcolsep}{2.5pt}
\renewcommand{\arraystretch}{0.86}
\resizebox{\columnwidth}{!}{%
\begin{tabular}{@{}lcccc@{}}
\toprule
Configuration & WV3 $33\times$ & WV3 $66\times$ & GF2 $50\times$ & Pavia $23.97\times$ \\
\midrule
Bands/$\latentdim$/$s$ & 8/8/4 & 8/16/8 & 4/16/8 & 102/64/4 \\
Stages/$d_0$ & 4/64 & 3/128 & 3/128 & 4/64 \\
Width multipliers & $[1,2,4,4]$ & $[2,4,4]$ & $[2,4,4]$ & $[1,2,4,4]$ \\
Encoder/decoder blocks & $[2,4,4,2]$/$[3,6,6,3]$ & $[2,4,2]$/$[2,4,2]$ & $[2,4,2]$/$[2,4,2]$ & $[2,4,4,2]$/$[2,4,4,2]$ \\
Embedding kernel & $64\!\times\!8\!\times\!2\!\times\!2$ & $128\!\times\!8\!\times\!2\!\times\!2$ & $128\!\times\!4\!\times\!2\!\times\!2$ & $64\!\times\!102\!\times\!2\!\times\!2$ \\
BSQ groups & 1 & 2 & 2 & 8 \\
Fusion blocks/input channels & 8/16 & 12/32 & 8/32 & 10/128 \\
\midrule
Stage-1 batch & 8 & 16 & 16 & 8 \\
Stage-1 MSE/BSQ weights & 1/0.05 & 4/0.1 & 4/0.1 & 1/0.1 \\
Fusion batch & 2 & 8 & 8 & 8 \\
\bottomrule
\end{tabular}}
\end{table}

The fusion module concatenates spectral and PAN codes, maps $2\latentdim$ channels to two logits per target bit, and uses 64-channel residual blocks with two $3\times3$ convolutions, residual addition, and GELU~\citep{Hendrycks2016GELU}. With the tokenizer frozen, it is trained to convergence using only Sec.~\ref{supp:app:bce_logits} (weight 1; image-domain losses disabled), seed 2025.

All tokenizers use EMA weights and three stages. Stage 1 jointly trains encoder and decoder for 40,000 iterations on mixed PAN, upsampled LR spectral, and HR reference images. Stages 2 and 3 initialize from the preceding EMA, freeze the encoder, and refine the decoder for 20,000 iterations each with batch 16 and MSE/Sobel-gradient/SAM weights $1/0.1/0.1$; quantizer loss is disabled. Stage 3 ramps bit-flip probability from $10^{-5}$ to $3\times10^{-3}$ over 2,000 iterations and holds it fixed. The stage-3 EMA is final. AdamW uses learning rate $2\times10^{-4}$, 500 tokenizer warm-up iterations, and weight decay 0.05/0.01 for tokenizer/fusion. No adversarial, perceptual, or LeCAM loss is used. Training uses PyTorch, bfloat16, and one NVIDIA RTX 4090.

Payload is $\latentdim/s^2$: WV3 $33\times$ uses $0.5$ bit/pixel ($0.0625$ bppb); WV3 $66\times$ and GF2 $50\times$ use $0.25$ bit/pixel ($0.03125$ and $0.0625$ bppb); Pavia uses $4$ bits/pixel, i.e., $64/(16\times102)=0.039216$ bppb and $23.9688\times$. WV3 $33\times$ sets \texttt{z\_channels}=512. Pavia pads to at most 128 channels while using 102 spectral and 64 latent channels.

\section{Datasets and Evaluation Protocols}
\label{supp:sec:supp_datasets}
WV3 (8 bands) and GF2 (4 bands) come from PanCollection~\citep{Deng2022MachineLearningPansharpening},\footnote{\url{https://github.com/liangjiandeng/PanCollection}} and the 102-band ROSIS Pavia Center data come from HyperPanCollection/HISI and Hyper-DSNet~\citep{zhuo2022deep}.\footnote{\url{https://github.com/liangjiandeng/HyperPanCollection}} All use scale factor four. Tab.~\ref{supp:tab:dataset_summary} gives every split used in training and evaluation; HR denotes the common spatial size of the reference or upsampled spectral image and PAN, and LR denotes the native spectral size.

\begin{table}[!t]
\centering
\caption{Dataset splits and spatial dimensions.}
\label{supp:tab:dataset_summary}
\label{supp:tab:dataset_wv3}
\label{supp:tab:dataset_gf2}
\label{supp:tab:dataset_pavia}
\scriptsize
\setlength{\tabcolsep}{3pt}
\renewcommand{\arraystretch}{0.92}
\begin{tabular*}{\columnwidth}{@{\extracolsep{\fill}}llrrrrc@{}}
\toprule
Dataset & Split & $N$ & $C$ & HR & LR & GT \\
\midrule
\multirow{4}{*}{WV3} & Train & 9714 & 8 & 64 & 16 & Yes \\
 & Val. & 1080 & 8 & 64 & 16 & Yes \\
 & Reduced & 20 & 8 & 256 & 64 & Yes \\
 & Full & 20 & 8 & 512 & 128 & No \\
\midrule
\multirow{4}{*}{GF2} & Train & 19809 & 4 & 64 & 16 & Yes \\
 & Val. & 2201 & 4 & 64 & 16 & Yes \\
 & Reduced & 20 & 4 & 256 & 64 & Yes \\
 & Full & 20 & 4 & 512 & 128 & No \\
\midrule
\multirow{3}{*}{Pavia} & Train & 1512 & 102 & 64 & 16 & Yes \\
 & Val. & 168 & 102 & 64 & 16 & Yes \\
 & Full & 2 & 102 & 400 & 100 & Yes \\
\bottomrule
\end{tabular*}
\end{table}

Arrays are channel-first $(N,C,H,W)$. The keys \texttt{gt}/\texttt{GT}, \texttt{lms}/\texttt{LMS}, \texttt{ms}/\texttt{MS}, and \texttt{pan}/\texttt{PAN} denote the HR reference, upsampled LR spectral image, native LR spectral image, and single-band PAN; WV3/GF2 use lowercase and Pavia uppercase. Communication accounting uses the native \texttt{ms}/\texttt{MS}. WV3/GF2 reduced-resolution tests follow Wald's protocol~\citep{Wald1997Protocol}; their GT supports PSNR, SSIM, ERGAS, and SAM, whereas full-resolution tests lack GT. Whenever HQNR, $D_{\lambda}$, and $D_s$ are reported, their anchors are the original spectral and PAN observations. Pavia \texttt{Full} retains GT, so both metric groups are available and are averaged over its two scenes. Its MAT variable \texttt{S} has shape $(400,400,102)$ and is permuted to $(102,400,400)$ before adding the sample dimension.

\section{Qualitative Comparisons With All Baselines Under JPEG2000}
\label{supp:sec:supp_jpeg2000_qualitative}
\begin{figure}[!t]
\centering
\includegraphics[width=\columnwidth,keepaspectratio]{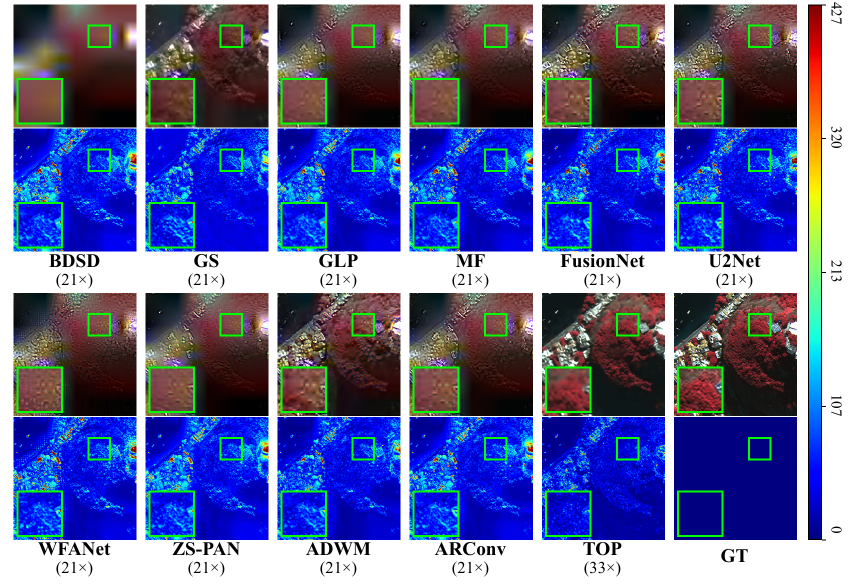}
\caption{WV3 reduced-resolution JPEG2000 comparison. Baselines: $21.26\times$; TOP: $33\times$.}
\label{supp:fig:wv3_qualitative_comparison}
\end{figure}

\begin{figure}[!t]
\centering
\includegraphics[width=\columnwidth,keepaspectratio]{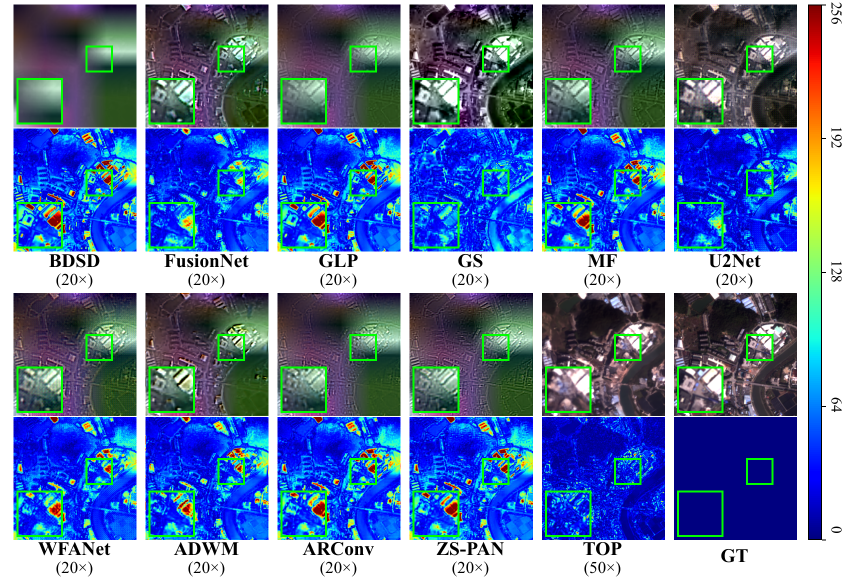}
\caption{GF2 reduced-resolution JPEG2000 comparison. Baselines: $20.33\times$; TOP: $50\times$.}
\label{supp:fig:gf2_qualitative_comparison}
\end{figure}

\begin{figure}[!t]
\centering
\includegraphics[width=\columnwidth,keepaspectratio]{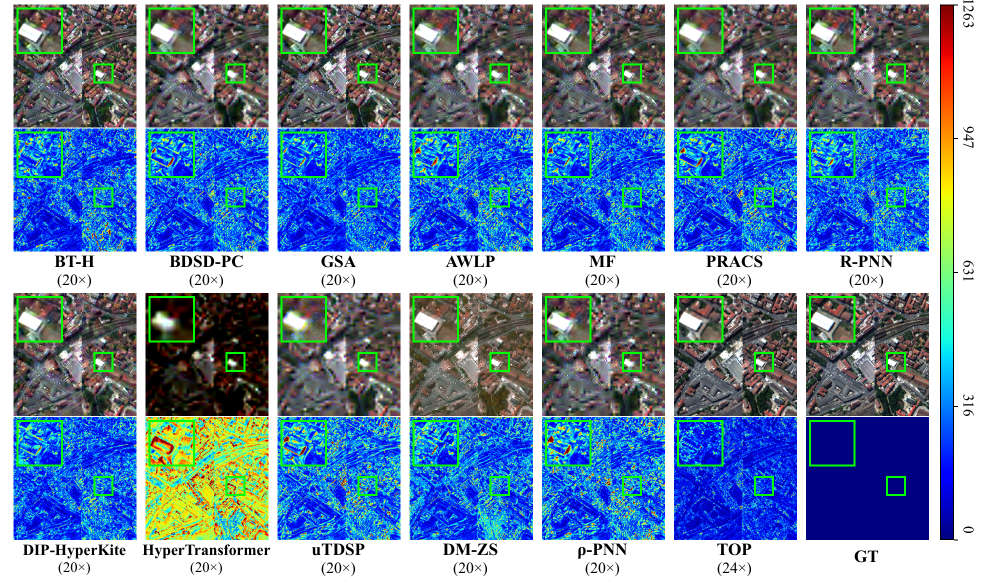}
\caption{Pavia \texttt{Full} JPEG2000 comparison. Baselines: $19.62\times$; TOP: $23.97\times$.}
\label{supp:fig:pavia_qualitative_comparison}
\end{figure}

For clarity, Fig.~\mainref{fig:qualitative_comparison} of the main manuscript shows only the best-performing baselines of each dataset. Figs.~\ref{supp:fig:wv3_qualitative_comparison}, \ref{supp:fig:gf2_qualitative_comparison}, and~\ref{supp:fig:pavia_qualitative_comparison} provide the comparisons with all baselines under the JPEG2000 protocol for the same WV3, GF2, and Pavia scenes.

\section{Comparisons Under the HEVC Protocol}
\label{supp:app:hevc_qualitative}
\begin{figure}[!t]
\centering
\includegraphics[width=\columnwidth,keepaspectratio]{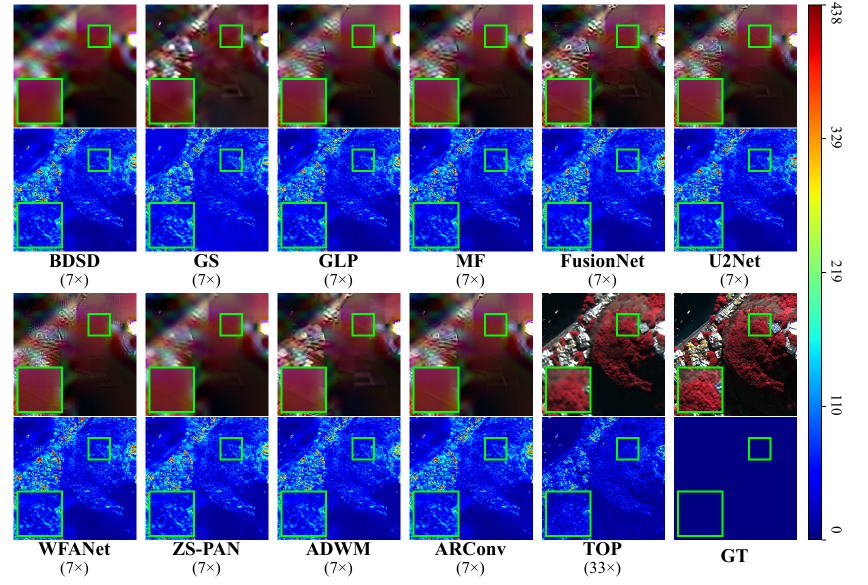}
\caption{WV3 reduced-resolution HEVC comparison. Baselines: $6.69\times$; TOP: $33\times$.}
\label{supp:fig:app_wv3_hevc_qualitative}
\end{figure}

\begin{figure}[!t]
\centering
\includegraphics[width=\columnwidth,keepaspectratio]{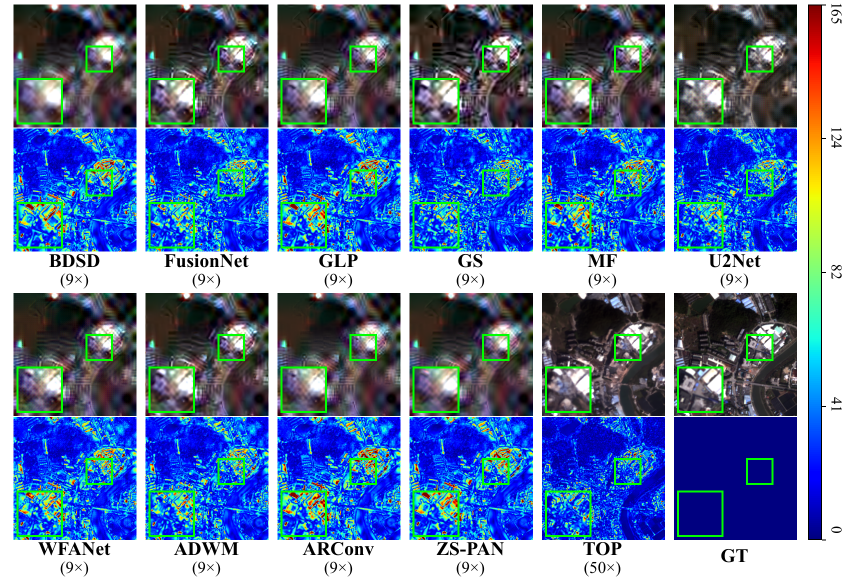}
\caption{GF2 reduced-resolution HEVC comparison. Baselines: $8.94\times$; TOP: $50\times$.}
\label{supp:fig:app_gf2_hevc_qualitative}
\end{figure}

\begin{figure}[!t]
\centering
\includegraphics[width=\columnwidth,keepaspectratio]{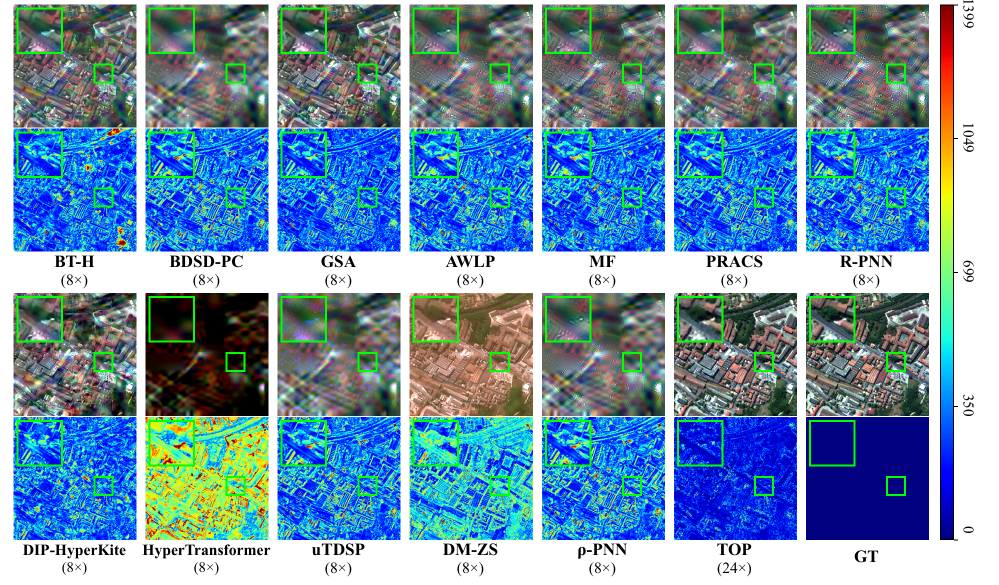}
\caption{Pavia \texttt{Full} HEVC comparison. Baselines: $7.79\times$ (displayed as $8\times$); TOP: $23.97\times$ (displayed as $24\times$).}
\label{supp:fig:app_pavia_hevc_qualitative}
\end{figure}

The main manuscript reports the reduced-resolution WV3 and GF2 HEVC results in Tabs.~\mainref{tab:main_wv3_hevc} and~\mainref{tab:main_gf2_hevc}. Tab.~\ref{supp:tab:app_pavia_hevc} gives the Pavia results, while Figs.~\ref{supp:fig:app_wv3_hevc_qualitative}--\ref{supp:fig:app_pavia_hevc_qualitative} provide the corresponding visual comparisons. The pixel-domain baselines operate at $6.69\times$, $8.94\times$, and $7.79\times$ on WV3, GF2, and Pavia, whereas TOP uses $33\times$, $50\times$, and $23.97\times$, respectively. Metric anchors follow Sec.~\ref{supp:sec:supp_datasets}; across all three datasets, TOP provides the strongest quantitative results and preserves clearer structures and more stable spectral appearance.


\begin{table*}[!htbp]
\caption{Pavia Full HEVC results averaged over two scenes.}
\label{supp:tab:app_pavia_hevc}
\centering
\scriptsize
\setlength{\tabcolsep}{3pt}
\renewcommand{\arraystretch}{0.90}
\begin{tabular*}{\textwidth}{@{\extracolsep{\fill}}lcccccccc@{}}
\toprule
Method & Ratio $\uparrow$ & PSNR $\uparrow$ & SSIM $\uparrow$ & ERGAS $\downarrow$ & HQNR $\uparrow$ & $D_\lambda$ $\downarrow$ & $D_s$ $\downarrow$ & SAM (deg) $\downarrow$ \\
\midrule
AWLP+HEVC  & \multirow{12}{*}{7.79$\times$} & $24.47$ & $0.57$ & $10.79$ & $0.67$ & $0.16$ & $0.20$ & $13.36$ \\
BDSD-PC+HEVC  &  & $24.28$ & $0.49$ & $11.05$ & $0.64$ & $0.16$ & $0.24$ & $12.32$ \\
BT-H+HEVC  &  & $25.11$ & $\underline{0.67}$ & $10.38$ & $0.60$ & $0.33$ & $0.11$ & $12.65$ \\
GSA+HEVC  &  & $\underline{26.64}$ & $0.64$ & $\underline{8.74}$ & $0.74$ & $0.17$ & $0.11$ & $\underline{12.22}$ \\
MF+HEVC  &  & $24.77$ & $0.59$ & $10.43$ & $0.70$ & $0.16$ & $0.16$ & $12.67$ \\
PRACS+HEVC  &  & $25.10$ & $0.56$ & $10.07$ & $0.73$ & $0.17$ & $0.12$ & $12.33$ \\
uTDSP+HEVC  &  & $24.50$ & $0.44$ & $11.45$ & $0.58$ & $0.16$ & $0.31$ & $13.88$ \\
DIP-HyperKite+HEVC  &  & $25.23$ & $0.61$ & $9.94$ & $\underline{0.77}$ & $0.17$ & $\underline{0.08}$ & $12.61$ \\
HyperTransformer+HEVC  &  & $18.71$ & $0.17$ & $21.71$ & $0.40$ & $0.19$ & $0.50$ & $35.16$ \\
R-PNN+HEVC  &  & $24.67$ & $0.57$ & $10.59$ & $0.71$ & $0.16$ & $0.16$ & $12.87$ \\
DM-ZS+HEVC  &  & $24.56$ & $0.64$ & $11.95$ & $0.70$ & $0.23$ & $0.09$ & $13.60$ \\
$\rho$-PNN+HEVC  &  & $23.86$ & $0.50$ & $11.63$ & $0.62$ & $\underline{0.16}$ & $0.27$ & $13.36$ \\
\midrule
TOP (ours) & \textbf{23.97$\times$} & $\mathbf{29.72}$ & $\mathbf{0.83}$ & $\mathbf{5.91}$ & $\mathbf{0.85}$ & $\mathbf{0.14}$ & $\mathbf{0.02}$ & $\mathbf{6.39}$ \\
\bottomrule
\end{tabular*}
\end{table*}

\section{Full-Resolution Validation on WV3 and GF2}
\label{supp:sec:supp_full_resolution_msi}
\begin{figure}[!t]
\centering
\includegraphics[width=\columnwidth,keepaspectratio]{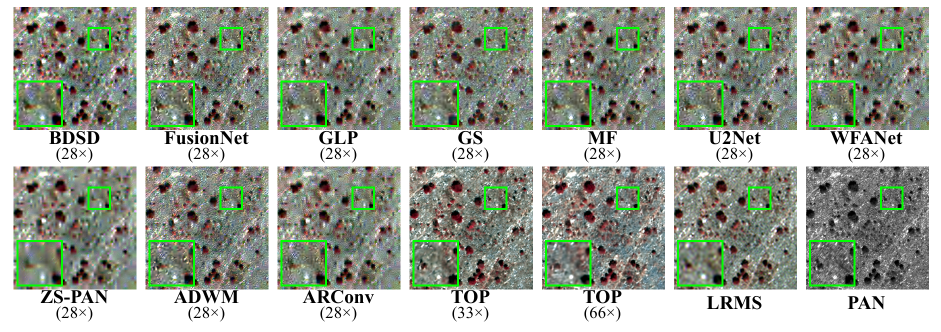}
\caption{WV3 full-resolution JPEG2000 comparison. Baselines: $28.43\times$; TOP: $33\times$ and $66\times$.}
\label{supp:fig:full_wv3_jpeg2000}
\end{figure}

\begin{figure}[!t]
\centering
\includegraphics[width=\columnwidth,keepaspectratio]{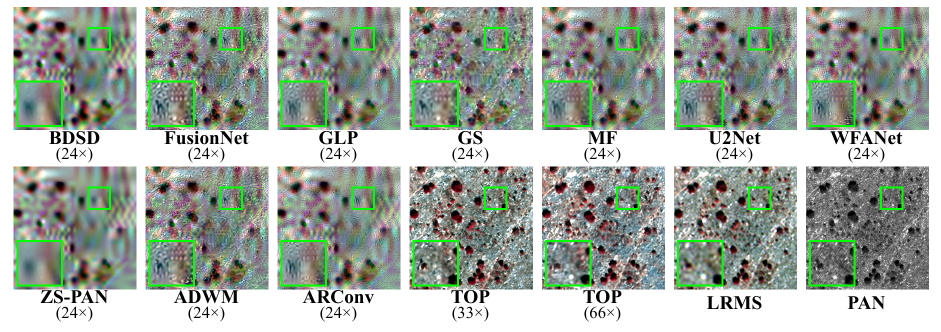}
\caption{WV3 full-resolution HEVC comparison. Baselines: $23.64\times$; TOP: $33\times$ and $66\times$.}
\label{supp:fig:full_wv3_hevc}
\end{figure}

\begin{figure}[!t]
\centering
\includegraphics[width=\columnwidth,keepaspectratio]{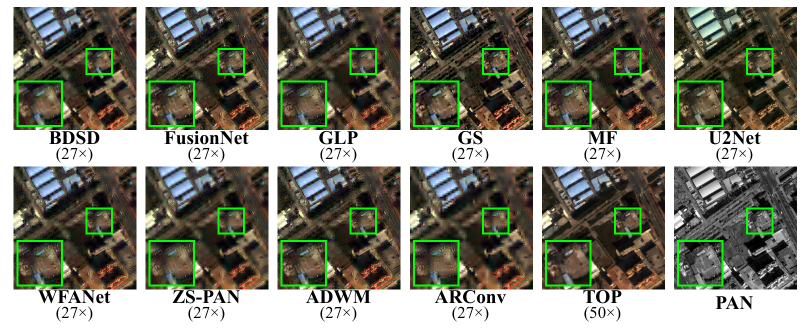}
\caption{GF2 full-resolution JPEG2000 comparison. Baselines: $27.46\times$; TOP: $50\times$.}
\label{supp:fig:full_gf2_jpeg2000}
\end{figure}

\begin{figure}[!t]
\centering
\includegraphics[width=\columnwidth,keepaspectratio]{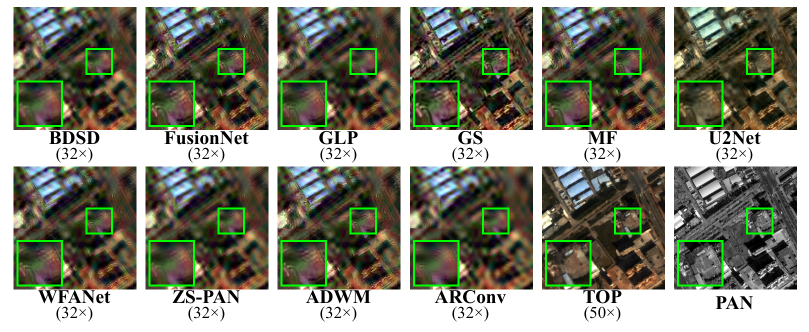}
\caption{GF2 full-resolution HEVC comparison. Baselines: $31.69\times$; TOP: $50\times$.}
\label{supp:fig:full_gf2_hevc}
\end{figure}

Full-resolution HR-MSI references are unavailable, so WV3/GF2 evaluation uses HQNR, $D_\lambda$, and $D_s$ against the original lossless PAN and LR-MSI. Pixel-domain baselines receive codec-decoded observations, TOP transmits its fused binary latent, and ZS-PAN follows its per-sample full-resolution protocol. Across both codecs and datasets, TOP attains the highest HQNR and lowest $D_\lambda$ at a higher ratio, and also the lowest $D_s$ under HEVC. Some JPEG2000 baselines have slightly lower $D_s$, but lower HQNR and weaker spectral consistency. Figs.~\ref{supp:fig:full_wv3_jpeg2000}--\ref{supp:fig:full_gf2_hevc} show the corresponding qualitative trend.

\begin{table}[!t]
\caption{WV3 full-resolution results under both codecs.}
\label{supp:tab:full_wv3_jpeg2000}
\label{supp:tab:full_wv3_hevc}
\centering
\scriptsize
\setlength{\tabcolsep}{2.5pt}
\renewcommand{\arraystretch}{0.82}
\resizebox{\columnwidth}{!}{%
\begin{tabular}{@{}lcccc@{}}
\toprule
Method & Ratio $\uparrow$ & HQNR $\uparrow$ & $D_\lambda$ $\downarrow$ & $D_s$ $\downarrow$ \\
\midrule
\multicolumn{5}{c}{JPEG2000} \\
\midrule
BDSD+JPEG2000 & \multirow{10}{*}{28.43$\times$} & $0.7925$ & $0.1636$ & $0.0524$ \\
FusionNet+JPEG2000 & & $0.7336$ & $0.1839$ & $0.1011$ \\
GLP+JPEG2000 & & $0.7342$ & $0.2043$ & $0.0769$ \\
GS+JPEG2000 & & $0.7983$ & $0.1756$ & $\mathbf{0.0322}$ \\
MF+JPEG2000 & & $0.7508$ & $0.1938$ & $0.0686$ \\
U2Net+JPEG2000 & & $0.7143$ & $0.1880$ & $0.1204$ \\
WFANet+JPEG2000 & & $0.6447$ & $0.2216$ & $0.1723$ \\
ADWM+JPEG2000 & & $0.7386$ & $0.1842$ & $0.0947$ \\
ARConv+JPEG2000 & & $0.6718$ & $0.2158$ & $0.1436$ \\
ZS-PAN+JPEG2000 & & $0.7082$ & $0.2049$ & $0.1101$ \\
\midrule
TOP (ours, $33\times$) & $\underline{33\times}$ & $\mathbf{0.8724}$ & $\mathbf{0.0984}$ & $\underline{0.0329}$ \\
TOP (ours, $66\times$) & $\mathbf{66\times}$ & $\underline{0.8481}$ & $\underline{0.1184}$ & $0.0389$ \\
\midrule
\multicolumn{5}{c}{HEVC} \\
\midrule
BDSD+HEVC & \multirow{10}{*}{23.64$\times$} & $0.3821$ & $0.4182$ & $0.3502$ \\
FusionNet+HEVC & & $0.3937$ & $0.4198$ & $0.3236$ \\
GLP+HEVC & & $0.3791$ & $0.4466$ & $0.3175$ \\
GS+HEVC & & $0.5420$ & $0.3300$ & $0.1939$ \\
MF+HEVC & & $0.4007$ & $0.4286$ & $0.3033$ \\
U2Net+HEVC & & $0.4108$ & $0.4066$ & $0.3111$ \\
WFANet+HEVC & & $0.3340$ & $0.4637$ & $0.3826$ \\
ADWM+HEVC & & $0.4383$ & $0.3835$ & $0.2900$ \\
ARConv+HEVC & & $0.3550$ & $0.4510$ & $0.3583$ \\
ZS-PAN+HEVC & & $0.3795$ & $0.4382$ & $0.3309$ \\
\midrule
TOP (ours, $33\times$) & $\underline{33\times}$ & $\mathbf{0.8724}$ & $\mathbf{0.0984}$ & $\mathbf{0.0329}$ \\
TOP (ours, $66\times$) & $\mathbf{66\times}$ & $\underline{0.8481}$ & $\underline{0.1184}$ & $\underline{0.0389}$ \\
\bottomrule
\end{tabular}}
\end{table}

\begin{table}[!t]
\caption{GF2 full-resolution results under both codecs.}
\label{supp:tab:full_gf2_jpeg2000}
\label{supp:tab:full_gf2_hevc}
\centering
\scriptsize
\setlength{\tabcolsep}{2.5pt}
\renewcommand{\arraystretch}{0.82}
\resizebox{\columnwidth}{!}{%
\begin{tabular}{@{}lcccc@{}}
\toprule
Method & Ratio $\uparrow$ & HQNR $\uparrow$ & $D_\lambda$ $\downarrow$ & $D_s$ $\downarrow$ \\
\midrule
\multicolumn{5}{c}{JPEG2000} \\
\midrule
BDSD+JPEG2000 & \multirow{10}{*}{27.46$\times$} & $0.8278$ & $0.1366$ & $0.0416$ \\
FusionNet+JPEG2000 & & $0.8304$ & $0.1366$ & $0.0382$ \\
GLP+JPEG2000 & & $0.7959$ & $0.1726$ & $0.0383$ \\
GS+JPEG2000 & & $0.7112$ & $0.1787$ & $0.1344$ \\
MF+JPEG2000 & & $0.7977$ & $0.1689$ & $0.0406$ \\
U2Net+JPEG2000 & & $0.7999$ & $\underline{0.1301}$ & $0.0800$ \\
WFANet+JPEG2000 & & $0.8150$ & $0.1535$ & $0.0374$ \\
ADWM+JPEG2000 & & $\underline{0.8311}$ & $0.1382$ & $\mathbf{0.0355}$ \\
ARConv+JPEG2000 & & $0.7298$ & $0.1846$ & $0.1048$ \\
ZS-PAN+JPEG2000 & & $0.8094$ & $0.1597$ & $\underline{0.0367}$ \\
\midrule
TOP (ours) & $\mathbf{50\times}$ & $\mathbf{0.8511}$ & $\mathbf{0.1031}$ & $0.0521$ \\
\midrule
\multicolumn{5}{c}{HEVC} \\
\midrule
BDSD+HEVC & \multirow{10}{*}{31.69$\times$} & $0.5165$ & $0.3248$ & $0.2365$ \\
FusionNet+HEVC & & $0.5704$ & $0.3048$ & $0.1801$ \\
GLP+HEVC & & $0.4497$ & $0.3841$ & $0.2709$ \\
GS+HEVC & & $0.6606$ & $0.2905$ & $\underline{0.0694}$ \\
MF+HEVC & & $0.4947$ & $0.3616$ & $0.2258$ \\
U2Net+HEVC & & $\underline{0.6819}$ & $\underline{0.2453}$ & $0.0971$ \\
WFANet+HEVC & & $0.4861$ & $0.3494$ & $0.2536$ \\
ADWM+HEVC & & $0.5499$ & $0.3164$ & $0.1964$ \\
ARConv+HEVC & & $0.4152$ & $0.3978$ & $0.3116$ \\
ZS-PAN+HEVC & & $0.4748$ & $0.3592$ & $0.2600$ \\
\midrule
TOP (ours) & $\mathbf{50\times}$ & $\mathbf{0.8511}$ & $\mathbf{0.1031}$ & $\mathbf{0.0521}$ \\
\bottomrule
\end{tabular}}
\end{table}

\begin{table*}[!t]
\centering
\caption{Onboard pansharpening followed by JPEG2000/HEVC on WV3. Full-reference metrics use GT; HQNR terms use the original LR-MSI/PAN. Dashes denote omitted CPU-only fusion times.}
\label{supp:tab:downlink_compression_combined}
\scriptsize
\renewcommand{\arraystretch}{0.88}
\resizebox{\textwidth}{!}{
\begin{tabular}{lccccccccc}
\toprule
Method & Ratio $\uparrow$ & Fusion time (ms/img) $\downarrow$ & HQNR $\uparrow$ & $D_\lambda$ $\downarrow$ & $D_s$ $\downarrow$ & PSNR $\uparrow$ & SSIM $\uparrow$ & ERGAS $\downarrow$ & SAM (deg) $\downarrow$ \\
\midrule
\multicolumn{10}{c}{\textbf{JPEG2000 Codec}} \\
\midrule
WFANet & 20.00$\times$ & 185.624 & $0.5847\pm0.0989$ & $0.2622\pm0.1046$ & $0.2103\pm0.0365$ & $\underline{29.2451}\pm3.9892$ & $\underline{0.6964}\pm0.1230$ & $\underline{7.1451}\pm1.4832$ & $8.0026\pm1.8765$ \\
U2Net & 20.03$\times$ & 114.427 & $0.5830\pm0.0999$ & $0.2658\pm0.1057$ & $0.2088\pm0.0367$ & $29.2254\pm3.9889$ & $0.6960\pm0.1230$ & $7.1554\pm1.4812$ & $8.0102\pm1.8692$ \\
FusionNet & 19.99$\times$ & \textbf{1.444} & $0.5824\pm0.0994$ & $0.2691\pm0.1064$ & $0.2057\pm0.0367$ & $29.2050\pm3.9294$ & $0.6961\pm0.1223$ & $7.1534\pm1.4757$ & $7.9752\pm1.9140$ \\
ADWM & 19.97$\times$ & 47.665 & $0.5794\pm0.0982$ & $0.2676\pm0.1035$ & $0.2116\pm0.0398$ & $29.1645\pm3.9383$ & $0.6947\pm0.1238$ & $7.1916\pm1.4841$ & $8.1475\pm1.8269$ \\
ARConv & 20.01$\times$ & 80.710 & $0.5879\pm0.1001$ & $0.2655\pm0.1065$ & $0.2024\pm0.0360$ & $29.1799\pm3.9359$ & $0.6952\pm0.1231$ & $7.1642\pm1.4738$ & $7.8361\pm1.8401$ \\
GS & 20.05$\times$ & - & $0.5464\pm0.0867$ & $0.2990\pm0.1064$ & $0.2197\pm0.0475$ & $28.4811\pm3.7461$ & $0.6658\pm0.1306$ & $7.7017\pm1.5986$ & $7.6836\pm1.9582$ \\
BDSD & 19.98$\times$ & - & $0.5857\pm0.0975$ & $0.2670\pm0.1130$ & $0.2020\pm0.0308$ & $28.6829\pm3.7466$ & $0.6830\pm0.1192$ & $7.5294\pm1.6008$ & $7.9723\pm1.8917$ \\
GLP & 19.97$\times$ & - & $0.6212\pm0.0939$ & $0.2483\pm0.1012$ & $\underline{0.1747}\pm0.0399$ & $28.9107\pm3.7803$ & $0.6869\pm0.1196$ & $7.3476\pm1.5406$ & $\underline{7.6171}\pm1.8623$ \\
MF & 20.02$\times$ & - & $\underline{0.6356}\pm0.0922$ & $\underline{0.2277}\pm0.0945$ & $0.1786\pm0.0435$ & $28.7851\pm3.7437$ & $0.6848\pm0.1226$ & $7.4207\pm1.5290$ & $7.6305\pm1.8427$ \\
ZS-PAN & \underline{20.07$\times$} & 36863.683 & $0.5629\pm0.1090$ & $0.3016\pm0.1267$ & $0.1942\pm0.0393$ & $28.5151\pm3.5605$ & $0.6838\pm0.1216$ & $7.4629\pm1.4911$ & $8.0665\pm1.8173$ \\
\midrule
TOP & \textbf{33$\times$} & \underline{32.511} & $\mathbf{0.7552}\pm0.1422$ & $\mathbf{0.1863}\pm0.1145$ & $\mathbf{0.0789}\pm0.0564$ & $\mathbf{30.3237}\pm2.8339$ & $\mathbf{0.8193}\pm0.0387$ & $\mathbf{6.0036}\pm1.2298$ & $\mathbf{5.9989}\pm1.3813$ \\
\midrule
\multicolumn{10}{c}{\textbf{HEVC Codec}} \\
\midrule
WFANet & 6.65$\times$ & 185.624 & $0.5355\pm0.1369$ & $0.3004\pm0.1321$ & $0.2437\pm0.0586$ & $28.6397\pm3.4463$ & $0.6983\pm0.0965$ & $7.3943\pm1.4707$ & $8.4511\pm1.8924$ \\
U2Net & 6.68$\times$ & 114.427 & $0.5267\pm0.1348$ & $0.3073\pm0.1322$ & $0.2484\pm0.0575$ & $28.5693\pm3.4832$ & $0.6952\pm0.0992$ & $7.4309\pm1.4827$ & $8.5067\pm1.9079$ \\
FusionNet & 6.66$\times$ & \textbf{1.444} & $0.5281\pm0.1415$ & $0.3115\pm0.1406$ & $0.2423\pm0.0577$ & $28.5932\pm3.4080$ & $0.6971\pm0.0965$ & $7.4091\pm1.4595$ & $8.4481\pm1.8467$ \\
ADWM & \underline{6.68$\times$} & 47.665 & $0.5232\pm0.1290$ & $0.3083\pm0.1269$ & $0.2515\pm0.0584$ & $28.5146\pm3.4834$ & $0.6923\pm0.1028$ & $7.4620\pm1.4281$ & $8.6357\pm1.8846$ \\
ARConv & 6.63$\times$ & 80.710 & $0.5424\pm0.1423$ & $0.3012\pm0.1366$ & $0.2332\pm0.0596$ & $\underline{28.6597}\pm3.3866$ & $\underline{0.6995}\pm0.0964$ & $\underline{7.3682}\pm1.4539$ & $8.2661\pm1.7866$ \\
GS & 6.65$\times$ & - & $0.5099\pm0.1286$ & $0.3311\pm0.1399$ & $0.2440\pm0.0530$ & $28.1831\pm3.4005$ & $0.6690\pm0.1047$ & $7.8659\pm1.6736$ & $\underline{8.0256}\pm1.9401$ \\
BDSD & 6.60$\times$ & - & $0.5598\pm0.1453$ & $0.2814\pm0.1393$ & $0.2311\pm0.0611$ & $28.4264\pm3.3264$ & $0.6910\pm0.0900$ & $7.6359\pm1.6836$ & $8.3336\pm1.8504$ \\
GLP & 6.60$\times$ & - & $0.5822\pm0.1529$ & $0.2744\pm0.1399$ & $\underline{0.2085}\pm0.0643$ & $28.6502\pm3.3278$ & $0.6950\pm0.0904$ & $7.4536\pm1.6026$ & $8.0500\pm1.8544$ \\
MF & 6.62$\times$ & - & $\underline{0.5861}\pm0.1475$ & $\underline{0.2635}\pm0.1339$ & $0.2146\pm0.0667$ & $28.4843\pm3.3915$ & $0.6895\pm0.0984$ & $7.5802\pm1.6303$ & $8.1932\pm1.8319$ \\
ZS-PAN & 6.66$\times$ & 36863.683 & $0.5327\pm0.1424$ & $0.3137\pm0.1384$ & $0.2329\pm0.0634$ & $28.2744\pm3.2000$ & $0.6888\pm0.0956$ & $7.6057\pm1.5630$ & $8.5449\pm1.8477$ \\
\midrule
TOP & \textbf{33$\times$} & \underline{32.511} & $\mathbf{0.7552}\pm0.1422$ & $\mathbf{0.1863}\pm0.1145$ & $\mathbf{0.0789}\pm0.0564$ & $\mathbf{30.3237}\pm2.8339$ & $\mathbf{0.8193}\pm0.0387$ & $\mathbf{6.0036}\pm1.2298$ & $\mathbf{5.9989}\pm1.3813$ \\
\bottomrule
\end{tabular}
}
\end{table*}

\section{Numerical Results of the Ablation Studies}
\label{supp:sec:supp_ablation}
All ablations use WV3 reduced-resolution data. Table~\ref{supp:tab:ablation_summary} reports (a) backbone, (b)--(c) BSQ payload/quality/complexity, (d) direct packing versus entropy model+rANS on the same latent, and (e) stage-3 robustness. The continuous-latent model is fully retrained with a larger fusion module; its entries use 32-bit latents. Both PSNR columns in (c) use the same 20 test scenes, unlike the validation value in Tab.~\mainref{tab:tobt_reconstruction}. In (e), $\Delta$PSNR is no-flip minus perturbed PSNR and $\Delta$SAM is perturbed minus no-flip SAM; lower is better.

\begin{table}[!t]
\caption{Numerical ablations on WV3 reduced-resolution data.}
\label{supp:tab:ablation_summary}
\centering
\scriptsize
\setlength{\tabcolsep}{2pt}
\renewcommand{\arraystretch}{0.82}
\resizebox{\columnwidth}{!}{%
\begin{tabular}{@{}lrrrr@{}}
\toprule
\multicolumn{5}{l}{\textbf{(a) Architecture reconstruction}} \\
Architecture & Params (M) $\downarrow$ & GFLOPs $\downarrow$ & PSNR $\pm$ Std. $\uparrow$ & SSIM $\pm$ Std. $\uparrow$ \\
\midrule
CNN & 66.721 & \textbf{169.96} & $\underline{27.42}\pm2.02$ & $\underline{0.75}\pm0.06$ \\
ViT & \textbf{62.709} & 219.46 & $27.05\pm1.82$ & $\underline{0.75}\pm0.06$ \\
Swin & \underline{65.609} & \underline{171.72} & $\mathbf{28.92}\pm2.07$ & $\mathbf{0.83}\pm0.04$ \\
\midrule
\multicolumn{5}{l}{\textbf{(b) BSQ payload}} \\
Method & bppb $\downarrow$ & Payload/latent entry $\downarrow$ & & \\
\midrule
TOBT & \textbf{0.0625} & \textbf{1 bit} & & \\
TOBT w/o BSQ & \underline{2.000} & \underline{32 bits} & & \\
\midrule
\multicolumn{5}{l}{\textbf{(c) BSQ reconstruction, fusion, and latent-module complexity}} \\
Method & Recon. PSNR $\uparrow$ & Fusion PSNR $\uparrow$ & Params (M) $\downarrow$ & GFLOPs $\downarrow$ \\
\midrule
TOBT & \underline{31.59} & \underline{30.32} & \textbf{0.61} & \textbf{0.31} \\
TOBT w/o BSQ & \textbf{34.70} & \textbf{32.11} & \underline{19.33} & \underline{9.88} \\
\midrule
\multicolumn{5}{l}{\textbf{(d) Direct packing versus entropy coding}} \\
Method & bppb $\downarrow$ & Add. params $\downarrow$ & Add. GFLOPs/sample $\downarrow$ & \\
\midrule
TOP & \underline{0.062500} & \textbf{0.00M} & \textbf{0.00} & \\
TOP$_{\mathrm{EM+rANS}}$ & \textbf{0.060260} & \underline{10.329096M} & \underline{74.769797} & \\
\midrule
\multicolumn{5}{l}{\textbf{(e) Bit-flip robustness on 20 scenes}} \\
Stage & Flip & $\Delta$PSNR (dB) $\downarrow$ & $\Delta$SAM (deg) $\downarrow$ & \\
\midrule
Stage 2 & 3\% & \underline{3.24} & \underline{0.36} & \\
Stage 3 & 3\% & \textbf{0.76} & \textbf{$-$0.11} & \\
Stage 2 & 5\% & \underline{5.17} & \underline{0.64} & \\
Stage 3 & 5\% & \textbf{1.61} & \textbf{$-$0.07} & \\
\bottomrule
\end{tabular}}
\end{table}

Panel (a) isolates the backbone choice. Swin improves reconstruction PSNR by 1.50~dB over CNN and 1.87~dB over ViT, while its 171.72~GFLOPs remain close to CNN and below ViT; the SSIM increase from 0.75 to 0.83 confirms better structural preservation. It therefore provides the strongest representation without using the most expensive candidate.

Panels (b)--(c) clarify the BSQ trade-off. Removing BSQ raises reconstruction and fusion PSNR, but changes each latent entry from one to 32 bits and requires about $31.7\times$ the fusion parameters and $31.9\times$ the GFLOPs. Its reconstruction--fusion gap also grows from 1.27 to 2.59~dB. BSQ therefore sacrifices some reconstruction ceiling for a target that is much smaller and easier for the onboard module to predict.

Panels (d)--(e) address the remaining communication risks. Entropy modeling saves only 3.58\% bppb while adding 10.33M parameters and 74.77~GFLOPs per sample, supporting direct packing. Stage-3 training reduces the PSNR loss under 3\% and 5\% flips by 76.5\% and 68.9\%, and leaves SAM nearly unchanged. The small negative $\Delta$SAM values reflect sample-level variation rather than a benefit from corruption; the meaningful result is the absence of systematic spectral degradation. The three components consequently have distinct roles: BSQ controls payload and prediction complexity, direct packing avoids disproportionate sender overhead, and decoder refinement limits residual bit errors.

\section{Comparison With Onboard Pansharpening Followed by Downlink Compression}
\label{supp:sec:supp_onboard_fusion}
This comparison tests whether moving conventional pansharpening on board is sufficient without redesigning the transmitted representation. Each baseline fuses the uncompressed PAN and LR-MSI into a full-resolution WV3 MSI and then applies JPEG2000 or HEVC; TOP predicts and bit-packs one fused latent. All methods use the same 20 scenes and ratio definition, so their payloads are measured against the same native observations.

Table~\ref{supp:tab:downlink_compression_combined} reports every quality metric and measured fusion time. Against JPEG2000, TOP improves the strongest baseline PSNR by 1.08~dB at $1.65\times$ its compression ratio, while raising the best HQNR from 0.6356 to 0.7552. Against HEVC, it improves the strongest PSNR by 1.66~dB at about $4.98\times$ the ratio. Its lower $D_{\lambda}$ and $D_s$ under both codecs show that the rate gain does not trade spectral consistency for spatial sharpness.

The payload difference follows from where compression occurs. After pixel-domain fusion, a baseline must represent all $CHW$ samples of the full-resolution MSI, and its final size depends on how strongly a generic codec can quantize that image. TOP instead transmits $\latentdim HW/s^2$ binary decisions, whose size is fixed before observing image content. For WV3 at $33\times$, this is 0.5 bit per spatial pixel, or 0.0625 bppb across eight bands. The comparison therefore evaluates two deployment representations, rather than only two fusion networks.

TOP has the same row in both codec blocks because it uses neither codec. The changing baseline rows show that onboard pixel-domain fusion still exposes the final product to a separately optimized lossy stage. The timing column measures fusion inference rather than complete encode--transmit--decode latency; among the reported neural baselines, only FusionNet is faster than TOP, while traditional CPU timings are omitted. Thus, the table supports the rate--quality advantage and moderate onboard fusion cost, but does not claim hardware-independent end-to-end latency or energy.

\section{Limitations}
\label{supp:sec:supp_limitations}

\paragraph{Sensor-Specific Training}
The band-adaptive embedding accepts different band counts, but dimensions and learned parameters remain configuration-specific. Each WV3, GF2, and Pavia setting in Tab.~\ref{supp:tab:implementation_schedule} therefore uses its own tokenizer and fusion module. A new sensor, or a ratio requiring a different latent configuration, requires one-time ground-side retraining without additional onboard computation. A sensor-agnostic tokenizer remains open.

\paragraph{Multi-Stage Training}
TOBT uses 40,000 joint encoder--decoder iterations, two 20,000-iteration decoder-refinement stages (the last with bit flips), and separate fusion training (Tab.~\ref{supp:tab:implementation_schedule}). These stages learn the latent space, adapt decoding to the transmitted HR code, and improve bit-error robustness, as supported by Sec.~\mainref{sec:exp_ablation_robustness}. The added training and tuning effort motivates a simpler future procedure.

\FloatBarrier

\end{document}